\documentclass[letterpaper,journal,onecolumn]{IEEEtran}
\usepackage{amsmath,amsfonts}
\usepackage{algorithmic}
\usepackage{algorithm}
\usepackage{array}
\usepackage[caption=false,font=normalsize,labelfont=sf,textfont=sf]{subfig}
\usepackage{textcomp}
\usepackage{stfloats}
\usepackage{url}
\usepackage{xurl}
\usepackage{cite}
\usepackage[T1]{fontenc}
\usepackage[latin9]{inputenc}
\usepackage{color}
\usepackage{cancel}
\usepackage[final,commentmarkup=footnote]{changes}
\setdeletedmarkup{\ifmmode\xcancel{#1}\else\sout{#1}\fi}
\definechangesauthor[name=Farzad, color=purple]{F}
\definechangesauthor[name=Claude, color=blue]{C}

\definecolor{document_fontcolor}{rgb}{0, 0, 0}
\definecolor{darkgreen}{rgb}{0.0, 0.5, 0.0}
\color{document_fontcolor}
\usepackage{units}
\usepackage{mathtools}
\mathtoolsset{showonlyrefs}
\usepackage{amsmath}
\usepackage{amsthm}
\usepackage{bbm}
\usepackage{amssymb}
\usepackage{dashbox}
\usepackage{graphicx}
\usepackage{esint}
\PassOptionsToPackage{normalem}{ulem}
\usepackage{ulem}
\usepackage{tikz}
\usepackage{hyperref}
\usetikzlibrary{decorations.pathreplacing,intersections}
\usetikzlibrary{fadings,decorations.text}

\usepackage{color}
\usetikzlibrary{patterns}
\usetikzlibrary{positioning}
\usepackage{csquotes}
\makeatletter

\theoremstyle{plain}
\newtheorem{thm}{\protect\theoremname}
\theoremstyle{plain}
\newtheorem{lem}{\protect\lemmaname}
\newtheorem{cor}{\protect\corollaryname}
\theoremstyle{definition}

\theoremstyle{definition}

\usepackage{thmtools,url}
\declaretheoremstyle[
spaceabove=3pt, spacebelow=3pt,
headfont=\normalfont\bfseries,
notefont=\mdseries, notebraces={(}{)},
bodyfont=\normalfont,
headformat=(\NAME\NUMBER),
postheadspace=.5em,
headpunct={}
]{mystyle}

\declaretheorem[style=mystyle,name=A,Refname=A,refname=A]{assumption}

\usepackage{mleftright}\mleftright
\date{}

\makeatother

\usepackage[english]{babel}
\addto\captionsenglish{\renewcommand{\figurename}{Fig.}}
\providecommand{\lemmaname}{Lemma}
\providecommand{\theoremname}{Theorem}
\providecommand{\remarkname}{Remark}
\providecommand{\observationname}{Observation}
\providecommand{\corollaryname}{Corollary}

\definecolor{ffc_teal}{rgb}{0.0,0.45,0.45}

\newcommand{\longversion}[1]{}

\newcommand{\cA}{\mathcal{A}}
\newcommand{\cB}{\mathcal{B}}
\newcommand{\cC}{\mathcal{C}}
\newcommand{\cD}{\mathcal{D}}
\newcommand{\cE}{\mathcal{E}}
\newcommand{\cF}{\mathcal{F}}
\newcommand{\cG}{\mathcal{G}}
\newcommand{\cH}{\mathcal{H}}
\newcommand{\cI}{\mathcal{I}}
\newcommand{\cJ}{\mathcal{J}}

\newcommand{\cL}{\mathcal{L}}

\newcommand{\cR}{\mathcal{R}}

\newcommand{\cX}{\mathcal{X}}

\newcommand{\bbE}{\mathbb{E}}

\newcommand{\bbP}{\mathbb{P}}

\newcommand{\bX}{{\boldsymbol{X}}}

\newcommand{\bh}{{\boldsymbol{h}}}

\newcommand{\bs}{{\boldsymbol{s}}}

\newcommand{\bu}{{\boldsymbol{u}}}
\newcommand{\bv}{{\boldsymbol{v}}}

\newcommand{\bx}{{\boldsymbol{x}}}

\begin{document}
\global\long\def\mult#1#2{\mu_{#1}^{#2}}
\global\long\def\mul#1{\boldsymbol{\mu}_{#1}}
\global\long\def\xmul#1{\boldsymbol{x}_{#1}}
\global\long\def\xmult#1#2{x_{#1}^{#2}}
\global\long\def\Mmul#1{\boldsymbol{M}_{#1}}
\global\long\def\hmul#1{\boldsymbol{h}_{#1}}
\global\long\def\hmult#1#2{h_{#1}^{#2}}
\global\long\def\diff#1#2{\delta_{#1}^{#2}}
\global\long\def\dif#1{\boldsymbol{\delta}_{#1}}
\global\long\def\muln#1{X_{#1}}
\global\long\def\ymul#1{\boldsymbol{y}_{#1}}
\global\long\def\ymult#1#2{y_{#1}^{#2}}
\global\long\def\zmul#1{\boldsymbol{z}_{#1}}
\global\long\def\zmult#1#2{z_{#1}^{#2}}
\global\long\def\rhomul#1{\boldsymbol{\rho}_{#1}}
\global\long\def\rhomult#1#2{\rho_{#1}^{#2}}
\global\long\def\init#1{\bar{x}^{#1}}
\global\long\def\ev{\mathbb{E}}
\global\long\def\ab{\mathcal{A}}
\global\long\def\eqlsyms#1#2{\left\langle #1,#2\right\rangle }
\global\long\def\nullsp{\operatorname{Null}}
\global\long\def\spn{\operatorname{Span}}
\global\long\def\maxr{\hat{m}}
\global\long\def\ga{\mathsf{A}}
\global\long\def\gc{\mathsf{C}}
\global\long\def\gg{\mathsf{G}}
\global\long\def\gt{\mathsf{T}}
\global\long\def\gx{\mathsf{X}}

\def\forsome{{\rm ~for~some~ }}
\def\forall{{\rm ~for~all~}}
\def\forany{{\rm ~for~any~}}
\def\th{{\rm th~}}
\def\RLL{{\rm RLL}}
\def\head{{\rm head}}
\def\tail{{\rm tail}}
\def\sur{{\rm sur}}
\def\Left{{\rm left}}
\def\Right{{\rm right}}

\newcommand{\abs}[1]{\left| #1 \right|}
\newcommand{\parenv}[1]{\left( #1 \right)}
\newcommand{\brckt}[1]{\left[ #1 \right]}
\newcommand{\mathset}[1]{\left\{ #1 \right\}}
\newcommand{\ccap}{\mathsf{cap}}
\newcommand{\floor}[1]{\left\lfloor #1\right\rfloor}
\newcommand{\ceil}[1]{\left\lceil #1\right\rceil}

\title{Asymptotic Analysis of Data Deduplication Under a Constant Edit Model}
\author{Sarvin~Motamen, Hao~Lou, and Farzad~Farnoud%
\thanks{This work was supported in part by the National Science Foundation under grants CIF-2312871 and CAREER-2144974. Earlier versions of this paper were presented in part at ISIT 2021 [DOI: \nolinkurl{10.1109/ISIT45174.2021.9517909}] and ISIT 2026 [DOI: \nolinkurl{10.1109/ISIT62367.2026.11653996}].}%
\thanks{S. Motamen is with the Department of Electrical and Computer Engineering, University of Virginia, Charlottesville, VA, USA (e-mail: sarvin@virginia.edu).}%
\thanks{H. Lou was with the Department of Electrical and Computer Engineering, University of Virginia, Charlottesville, VA, USA (e-mail: haolou@virginia.edu).}%
\thanks{Farzad Farnoud is with the Department of Electrical and Computer Engineering and the Department of Computer Science, University of Virginia, Charlottesville, VA 22903 USA (e-mail: farzad@virginia.edu).}}

\maketitle

\begin{abstract} 
    Data deduplication has become increasingly important in large-scale storage systems, driven by the rapid growth of digital data. By identifying and removing repeats in the data stream, deduplication saves storage space more efficiently than traditional compression methods. In this paper, we study the performance of variable-length deduplication and multi-chunk deduplication algorithms from an information-theoretic perspective, focusing on data streams where repeats are approximate rather than exact. We consider two probabilistic source models in which source strings consist of repeated blocks that are modified by either a constant number of substitution edits or burst-insertion edits per block, the latter capturing the localized edit patterns commonly observed in real-world data. For both models, we show that variable-length deduplication achieves asymptotically arbitrarily large compression ratios, and that multi-chunk deduplication achieves compression within a constant factor of the source entropy under mild conditions.
\end{abstract}

\begin{IEEEkeywords}
Data deduplication, data compression, source coding, substitution edits, burst insertions.
\end{IEEEkeywords}

\section{Introduction}
With the explosive growth of digital data, especially redundant data~\cite{gantz2012digital,meyer2012study,el2012primary}, the challenge of reducing data storage costs is drawing increasing interest.
Data deduplication was proposed to identify and eliminate duplicates in input data streams or files, thus reducing storage space.
This goal is achieved by parsing the data stream into multiple data ``chunks'', where each chunk is entered into a dictionary upon first occurrence, and replaced by pointers to the dictionary for all duplicates.
Data deduplication is computationally more efficient than traditional data compression methods, which has led to its widespread adoption in mass data storage systems, e.g., LBFS~\cite{muthitacharoen2001low} and Venti~\cite{quinlan2002venti}.

Data deduplication algorithms divide data into chunks using either fixed-length or content-defined variable-length schemes. Fixed-Length Deduplication (FLD) is simple but suffers from the boundary-shift problem: insertions or deletions shift all subsequent chunks, reducing effectiveness. In contrast, Variable-Length Deduplication (VLD) and Multi-Chunk Deduplication (MCD), introduced in~\cite{niesen2019information} and studied in this paper, use content-defined chunking (CDC), where a sliding window sets chunk boundaries based on predefined patterns called \emph{markers}. This makes chunking robust to local edits. VLD and MCD are described in detail in Section~\ref{sec:alg}. CDC is widely used in practical deduplication systems, e.g.,~\cite{muthitacharoen2001low,el2012primary, bobbarjung2006fingerdiff,teodosiu2006optimizing,agarwal2010endre,zhang2015ae,yu2015leap,kruus2010bimodal,lu2010frequency,zhou2013hysteresis}.
In CDC, a substantial fraction of the chunks can be extremely small or large. With large chunk sizes, duplicates in data tend to remain undetected, leading to ineffective deduplication. On the other hand, small chunks increase VLD's metadata overhead through dictionary pointers and indicator bits that distinguish new chunks from repeated ones. MCD can be regarded as a modification of VLD to address the problem caused by small chunks.

While data deduplication has been extensively studied from a practical standpoint (see~\cite{xia2016comprehensive} for a survey), its theoretical analysis from an information-theoretic perspective remains limited, despite the fact that, as a form of data compression, data deduplication is inherently an information-theoretic problem with fundamental lower bounds on performance.
The first information-theoretic analysis of deduplication was presented in the seminal work of Niesen~\cite{niesen2019information}, where VLD and MCD were formalized and studied. Specifically, in~\cite{niesen2019information}, the performance of certain deduplication algorithms was studied using a probabilistic model for a data stream constructed as follows. The data stream $\bs$ is composed of $B$ blocks, each being an exact copy of a source symbol, where the source symbols belong to an alphabet consisting of $A$ random binary strings of average length $L$ bits. The effect of the marker length $M$ on the performance of VLD and MCD was studied, showing that VLD can achieve arbitrarily large compression ratios (expected input length divided by expected compressed length), but may be polynomially suboptimal relative to the source entropy. MCD was shown to be order optimal under mild conditions, meaning that its expected compressed length is within a constant factor of the source entropy.

While~\cite{niesen2019information} considers blocks of data to be exactly repeated in the data stream, in real-world scenarios, copies of repeated data blocks are often approximate rather than exact, due to edits to the data. With this motivation, VLD was studied in~\cite{lou2022data} over a probabilistic source model where edits are modeled as random substitution, with each bit independently flipped with a given probability. Specifically, rather than being an exact copy, each of the $B$ blocks is generated by taking a source symbol and flipping each bit with a constant probability $\delta$.
The performance of VLD was characterized in terms of $\delta$ and the marker length $M$. Under the asymptotic assumptions of that work, with $0<\delta<1/2$ fixed as $B\to\infty$, the compression ratio is upper bounded by a universal constant if $2^M=o\left(\log B\right)$ or $2^M=\omega\left(\log B\right)$. This holds even for arbitrarily small fixed $\delta$, although the limiting source entropy per input bit tends to zero as $\delta\to0$. On the other hand, with the optimal choice $2^M = \Theta\left(\log\left(B/A\right)\right)$, VLD achieves arbitrarily large compression ratios as $\delta \to 0$. However, the upper bound on the ratio of the length of the compressed sequence to entropy grows as $\delta$ decreases, suggesting that the gap to entropy may increase for smaller edit probabilities; whether this gap is fundamental or an artifact of the analysis remains an open question.
Deduplication with approximate repeats was also analyzed in~\cite{vestergaard2019generalized} from an information-theoretic point of view but with a different source model and algorithm. In their model, all source blocks have equal length and undergo a limited number of random substitution edits.
We note that the problem of deduplication under edits was also considered in~\cite{conde2018data}, which focused on performing deduplication on two files, one being an edited version of the other by insertions and deletions, where the edits are random and not localized. In this work, we focus on deduplicating one data stream, where symbols of various lengths are repeated.

While the analysis in \cite{lou2022data} enhances our understanding of deduplication under edits and characterizes the dependence of the optimal marker length on the prevalence of edits, the edit model used, i.e., randomly distributed substitutions, does not fit real-world data, where edits that appear close to each other, referred to as burst or localized edits, are common in a wide range of applications, including file synchronization.
The prevalence of burst edits is demonstrated in~\cite{tang2025burst} by statistical analysis of real data, specifically, two versions of the Bash source code. The study found that the hypothesis of uniform edits was rejected for over 97\% of the files in the dataset, using a p-value threshold of 5\%. It is thus of interest to study deduplication under a burst edit model, deriving upper and lower bounds on the algorithms' performance and optimizing the parameters of the algorithms as a function of model parameters.


Beyond the choice of edit model, the bounds derived in~\cite{lou2022data} have another limitation. The analysis holds the edit probability fixed as the data length grows, so the expected number of edits grows linearly with the data length and the bounds do not directly characterize sparsely edited data. This also leads to a gap in our understanding, where the regimes with no edits and linear number of edits are studied, but the sublinear case that bridges them is not. When $AL+B\log A=o\left(BL\right)$, the no-edit model has entropy $O\left(AL+B\log A\right)=o\left(BL\right)$. In contrast, the independent-substitution model with fixed $0<\delta<1/2$ has entropy $\Theta\left(BL\right)$~\cite[Lemma~2]{lou2022data}. Deduplication in the first case can achieve compressed files whose length is of a different order than the uncompressed file, while in the second case, the compressed file is of the same order as the uncompressed file. It is thus of interest to analyze the regime where each source block undergoes a constant number of edits, which can be regarded as a bridge between the two regimes, and understand how the performance of deduplication algorithms changes as we transition from one regime to the other.

Lou and Farnoud~\cite{lou2022data} analyze deduplication under independent substitution edits, with the edit probability $\delta$ held fixed as the data length grows.
Their results show that appropriately choosing the marker length allows VLD to achieve arbitrarily large compression ratios as the fixed edit probability becomes small.
However, these fixed-$\delta$ asymptotic results do not directly characterize the sparse-edit regime $\delta=\Theta\left(1/L\right)$, where the expected number of substitutions per block remains constant.
This motivates studying sparse edits directly through source models in which each block undergoes exactly $t$ edits, with $t$ fixed independently of its length.
We consider two such models: constant-number single-bit substitutions and constant-number burst insertions, each of length $k$.

We focus on substitutions in the first model and insertions in the second rather than all edit types for the sake of simplicity.
Under the respective conditions of Corollaries~\ref{cor:vldsub_summary} and~\ref{cor:vldins_summary}, we show that VLD with optimal marker length can have expected compressed length of greater order than the source entropy under both edit models while still achieving asymptotically arbitrarily large compression ratios. In the burst-insertion model, under these conditions, this gap occurs when $k=o\left(\sqrt{L\log B}\right)$, whereas VLD is order optimal when $k=\Omega\left(\sqrt{L\log B}\right)$. Further, under mild conditions, MCD is shown to be order-optimal for both source models.



This paper extends our conference studies~\cite{lou2021asymptotic,motamen2026burst}, which analyzed VLD and MCD under constant substitutions and VLD under burst insertions, respectively. For constant substitutions, we derive a family of VLD lower bounds that accounts for the relationship between the alphabet size, source-symbol length, and number of source blocks. For burst insertions, we extend the analysis to MCD and establish its order optimality under the stated conditions. We also strengthen the VLD lower bounds to account explicitly for the inserted bits, obtaining matching asymptotic bounds under the conditions of Corollary~\ref{cor:vldins_summary} and identifying the insertion-length regimes in which VLD is order optimal. Finally, we provide complete proofs for results whose proofs were omitted or sketched in the conference papers.

The rest of the paper is organized as follows. Notation and preliminaries are given in the next section. In Section~\ref{sec:source}, we describe the studied information source models and bound their entropies. In Section~\ref{sec:alg}, we formally describe the deduplication algorithms VLD and MCD that are analyzed in the rest of the paper. Bounds on the performance of the algorithms are derived in Sections~\ref{sec:vld} and~\ref{sec:mcd}. In Section~\ref{sec:results}, we summarize the main results of this paper. Finally, we conclude the paper in Section~\ref{sec:conclusion}.

\section{Notation and Preliminaries}\label{sec:pre}
All logarithms are to base 2. Binary strings are denoted by boldface lowercase letters, such as $\bx$ and $\bu$, or by explicitly defined uppercase notation for particular string-valued objects, such as source symbols, source blocks, and chunks. Individual bits are denoted by plain lowercase letters or by their values $0$ and $1$. The term \emph{source symbol} refers to an entire binary string, rather than an individual bit. 

We consider the binary alphabet $\{0,1\}$, denoted $\Sigma$. The set of all finite strings over $\Sigma$ is denoted $\Sigma^*$. For a positive integer $m$, let $\Sigma^{m}$ be the set of all strings of length $m$ over $\Sigma$. For strings $\bu,\bv\in\Sigma^*$, the concatenation of $\bu$ and $\bv$ is denoted $\bu\bv$. The length-$\ell$ substring of $\bu$ starting from the $i$-th symbol is denoted $\bu_{i,\ell}$. The length of $\bu$ is denoted $\abs{\bu}$. The cardinality of a set $S$ is also denoted $\abs{S}$. The Hamming distance between $\bu$ and $\bv$ is denoted $d\left(\bu,\bv\right)$.

For any $0\le x\le 1$, $H\left(x\right)$ denotes the binary entropy function, i.e., $H\left(x\right)=x\log \left(\frac{1}{x}\right)+\left(1-x\right)\log \left(\frac{1}{1-x}\right)$. For any event $\cE$, $\bar\cE$ denotes the complement of $\cE$, and we define the indicator variable $\mathbbm{1}_{\cE}$ to be 1 if $\cE$ is true, and 0 otherwise. The $k$-runlength-limited (RLL) strings~\cite{marcus2001introduction} are binary strings that do not contain $k$ consecutive zeros. For any non-negative integer $n$, we denote the set of all $k$-RLL strings of length $n$ by $R_k^n$.



\section{Source models}\label{sec:source}
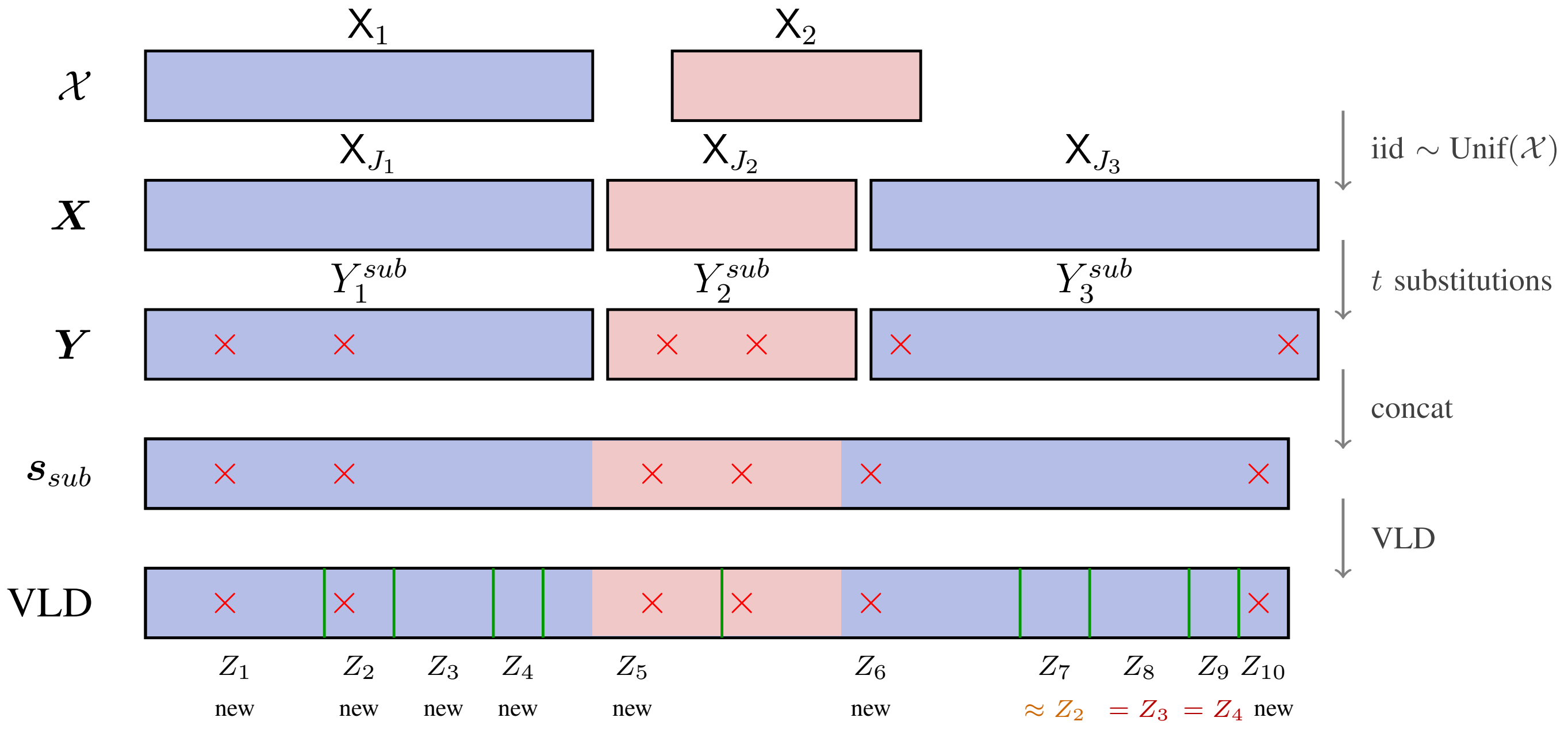
\begin{figure*}[t]
    \centering
    \resizebox{\textwidth}{!}{%
    \begin{tikzpicture}[
        thick,
        lbl/.style={font=\large, anchor=east},
        xlbl/.style={font=\large\bfseries, inner sep=1pt},
        xmark/.style={red, font=\large\bfseries},
    ]

    \definecolor{col1}{RGB}{180,190,230}   
    \definecolor{col2}{RGB}{240,200,200}   

    \def\lenA{4.5}    
    \def\lenB{2.5}    
    \def\blkh{0.7}    
    \def\rsep{1.3}    
    \def\lblx{-0.4}   
    \def\laboff{0.02}  

    \def\rowy{0}
    \node[lbl] at (\lblx, \rowy) {$\mathcal{X}$};

    \draw[thick, fill=col1] (0, \rowy-\blkh/2) rectangle (\lenA, \rowy+\blkh/2);
    \node[xlbl, anchor=south] at (\lenA/2, \rowy+\blkh/2+\laboff) {$\gx_1$};

    \pgfmathsetmacro{\xbstart}{\lenA + 0.8}
    \draw[thick, fill=col2] (\xbstart, \rowy-\blkh/2) rectangle (\xbstart+\lenB, \rowy+\blkh/2);
    \node[xlbl, anchor=south] at (\xbstart+\lenB/2, \rowy+\blkh/2+\laboff) {$\gx_2$};

    \pgfmathsetmacro{\rowy}{-\rsep}
    \node[lbl] at (\lblx, \rowy) {$\boldsymbol{X}$};

    \draw[thick, fill=col1] (0, \rowy-\blkh/2) rectangle (\lenA, \rowy+\blkh/2);
    \node[xlbl, anchor=south] at (\lenA/2, \rowy+\blkh/2+\laboff) {$\gx_{J_1}$};

    \pgfmathsetmacro{\jbstart}{\lenA + 0.15}
    \draw[thick, fill=col2] (\jbstart, \rowy-\blkh/2) rectangle (\jbstart+\lenB, \rowy+\blkh/2);
    \node[xlbl, anchor=south] at (\jbstart+\lenB/2, \rowy+\blkh/2+\laboff) {$\gx_{J_2}$};

    \pgfmathsetmacro{\jcstart}{\jbstart + \lenB + 0.15}
    \pgfmathsetmacro{\annx}{\jcstart + \lenA + 0.4}
    \draw[thick, fill=col1] (\jcstart, \rowy-\blkh/2) rectangle (\jcstart+\lenA, \rowy+\blkh/2);
    \node[xlbl, anchor=south] at (\jcstart+\lenA/2, \rowy+\blkh/2+\laboff) {$\gx_{J_3}$};

    \pgfmathsetmacro{\rowy}{-2*\rsep}
    \node[lbl] at (\lblx, \rowy) {$\boldsymbol{Y}$};

    \draw[thick, fill=col1] (0, \rowy-\blkh/2) rectangle (\lenA, \rowy+\blkh/2);
    \node[xlbl, anchor=south] at (\lenA/2, \rowy+\blkh/2+\laboff) {$Y^{sub}_1$};
    \node[xmark] at (0.8, \rowy) {$\times$};
    \node[xmark] at (2.0, \rowy) {$\times$};

    \draw[thick, fill=col2] (\jbstart, \rowy-\blkh/2) rectangle (\jbstart+\lenB, \rowy+\blkh/2);
    \node[xlbl, anchor=south] at (\jbstart+\lenB/2, \rowy+\blkh/2+\laboff) {$Y^{sub}_2$};
    \node[xmark] at (\jbstart+0.6, \rowy) {$\times$};
    \node[xmark] at (\jbstart+1.5, \rowy) {$\times$};

    \draw[thick, fill=col1] (\jcstart, \rowy-\blkh/2) rectangle (\jcstart+\lenA, \rowy+\blkh/2);
    \node[xlbl, anchor=south] at (\jcstart+\lenA/2, \rowy+\blkh/2+\laboff) {$Y^{sub}_3$};
    \node[xmark] at (\jcstart+0.3, \rowy) {$\times$};
    \node[xmark] at (\jcstart+4.2, \rowy) {$\times$};

    \pgfmathsetmacro{\totalw}{\lenA + \lenB + \lenA}

    \node[font=\small, black!75, anchor=west] at (\annx, -5*\rsep/2) {concat};
    \draw[->, gray] (\annx - 0.15, -5*\rsep/2 + 0.4) -- (\annx - 0.15, -5*\rsep/2 - 0.4);

    \pgfmathsetmacro{\rowy}{-3*\rsep}
    \node[lbl] at (\lblx, \rowy) {$\bs_{sub}$};

    \draw[thick, fill=col1] (0, \rowy-\blkh/2) rectangle (\totalw, \rowy+\blkh/2);
    \fill[col2] (\lenA, \rowy-\blkh/2+0.02) rectangle (\lenA+\lenB, \rowy+\blkh/2-0.02);
    \fill[col1] (\lenA+\lenB, \rowy-\blkh/2+0.02) rectangle (\totalw-0.02, \rowy+\blkh/2-0.02);
    \draw[thick] (0, \rowy-\blkh/2) rectangle (\totalw, \rowy+\blkh/2);

    \node[xmark] at (0.8, \rowy) {$\times$};
    \node[xmark] at (2.0, \rowy) {$\times$};
    \node[xmark] at (\lenA+0.6, \rowy) {$\times$};
    \node[xmark] at (\lenA+1.5, \rowy) {$\times$};
    \node[xmark] at (\lenA+\lenB+0.3, \rowy) {$\times$};
    \node[xmark] at (\lenA+\lenB+4.2, \rowy) {$\times$};

    \node[font=\small, black!75, anchor=west] at (\annx, -7*\rsep/2) {VLD};
    \draw[->, gray] (\annx - 0.15, -7*\rsep/2 + 0.4) -- (\annx - 0.15, -7*\rsep/2 - 0.4);

    \pgfmathsetmacro{\rowy}{-4*\rsep}
    \pgfmathsetmacro{\yThreeStart}{\lenA+\lenB}

    \draw[thick, fill=col1] (0, \rowy-\blkh/2) rectangle (\totalw, \rowy+\blkh/2);
    \fill[col2] (\lenA, \rowy-\blkh/2+0.02) rectangle (\lenA+\lenB, \rowy+\blkh/2-0.02);
    \fill[col1] (\yThreeStart, \rowy-\blkh/2+0.02) rectangle (\totalw-0.02, \rowy+\blkh/2-0.02);
    \draw[thick] (0, \rowy-\blkh/2) rectangle (\totalw, \rowy+\blkh/2);

    \node[xmark] at (0.8, \rowy) {$\times$};
    \node[xmark] at (2.0, \rowy) {$\times$};
    \node[xmark] at (\lenA+0.6, \rowy) {$\times$};
    \node[xmark] at (\lenA+1.5, \rowy) {$\times$};
    \node[xmark] at (\yThreeStart+0.3, \rowy) {$\times$};
    \node[xmark] at (\yThreeStart+4.2, \rowy) {$\times$};

    \pgfmathsetmacro{\mkA}{1.8}              
    \pgfmathsetmacro{\mkAb}{2.5}             
    \pgfmathsetmacro{\mkB}{3.5}              
    \pgfmathsetmacro{\mkC}{4.0}              
    \pgfmathsetmacro{\mkD}{5.8}              
    \pgfmathsetmacro{\mkE}{\yThreeStart+1.8} 
    \pgfmathsetmacro{\mkEb}{\yThreeStart+2.5}
    \pgfmathsetmacro{\mkF}{\yThreeStart+3.5} 
    \pgfmathsetmacro{\mkG}{\yThreeStart+4.0} 
    \foreach \mx in {\mkA, \mkAb, \mkB, \mkC, \mkD, \mkE, \mkEb, \mkF, \mkG} {
        \draw[green!60!black, thick] (\mx, \rowy-\blkh/2) -- (\mx, \rowy+\blkh/2);
    }

    \node[lbl] at (\lblx, \rowy) {VLD};

    \pgfmathsetmacro{\cA}{\mkA/2}
    \pgfmathsetmacro{\cB}{(\mkA+\mkAb)/2}
    \pgfmathsetmacro{\cC}{(\mkAb+\mkB)/2}
    \pgfmathsetmacro{\cD}{(\mkB+\mkC)/2}
    \pgfmathsetmacro{\cE}{(\mkC+\mkD)/2}
    \pgfmathsetmacro{\cF}{(\mkD+\mkE)/2}
    \pgfmathsetmacro{\cG}{(\mkE+\mkEb)/2}
    \pgfmathsetmacro{\cH}{(\mkEb+\mkF)/2}
    \pgfmathsetmacro{\cI}{(\mkF+\mkG)/2}
    \pgfmathsetmacro{\cJ}{(\mkG+\totalw)/2}
    \node[font=\footnotesize, below] at (\cA, \rowy-\blkh/2-0.05) {$Z_1$};
    \node[font=\footnotesize, below] at (\cB, \rowy-\blkh/2-0.05) {$Z_2$};
    \node[font=\footnotesize, below] at (\cC, \rowy-\blkh/2-0.05) {$Z_3$};
    \node[font=\footnotesize, below] at (\cD, \rowy-\blkh/2-0.05) {$Z_4$};
    \node[font=\footnotesize, below] at (\cE, \rowy-\blkh/2-0.05) {$Z_5$};
    \node[font=\footnotesize, below] at (\cF, \rowy-\blkh/2-0.05) {$Z_6$};
    \node[font=\footnotesize, below] at (\cG, \rowy-\blkh/2-0.05) {$Z_7$};
    \node[font=\footnotesize, below] at (\cH, \rowy-\blkh/2-0.05) {$Z_8$};
    \node[font=\footnotesize, below] at (\cI, \rowy-\blkh/2-0.05) {$Z_9$};
    \node[font=\footnotesize, below] at (\cJ, \rowy-\blkh/2-0.05) {$Z_{10}$};

    \pgfmathsetmacro{\elaby}{\rowy-\blkh/2-0.73}
    \node[font=\scriptsize] at (\cA, \elaby) {new};
    \node[font=\scriptsize] at (\cB, \elaby) {new};
    \node[font=\scriptsize] at (\cC, \elaby) {new};
    \node[font=\scriptsize] at (\cD, \elaby) {new};
    \node[font=\scriptsize] at (\cE, \elaby) {new};
    \node[font=\scriptsize] at (\cF, \elaby) {new};
    \node[font=\scriptsize\bfseries, orange!80!black] at (\cG, \elaby) {$\approx Z_2$};
    \node[font=\scriptsize\bfseries, red!70!black] at (\cH, \elaby) {$=Z_3$};
    \node[font=\scriptsize\bfseries, red!70!black] at (\cI, \elaby) {$=Z_4$};
    \node[font=\scriptsize, xshift=3pt] at (\cJ, \elaby) {new};

    \node[font=\small, black!75, anchor=west] at (\annx, -\rsep/2) {iid $\sim\text{Unif}\left(\mathcal{X}\right)$};
    \draw[->, gray] (\annx - 0.15, -\rsep/2 + 0.4) -- (\annx - 0.15, -\rsep/2 - 0.4);

    \node[font=\small, black!75, anchor=west] at (\annx, -3*\rsep/2) {$t$ substitutions};
    \draw[->, gray] (\annx - 0.15, -3*\rsep/2 + 0.4) -- (\annx - 0.15, -3*\rsep/2 - 0.4);

    \end{tikzpicture}%
    }
    \caption{Illustration of the constant-substitution source model with $A=2$ source symbols and $B=3$ source blocks, where $J_1=J_3=1$ and $J_2=2$, and with $t=2$ substitutions per block. Here $\boldsymbol X=\left(\gx_{J_1},\ldots,\gx_{J_B}\right)$ and $\boldsymbol Y=\left(Y^{sub}_1,\ldots,Y^{sub}_B\right)$. The bottom row shows the VLD chunking of $\bs_{sub}$ by the marker $0^M$, producing chunks $Z_1,\ldots,Z_{10}$. Chunks $Z_8$ and $Z_9$ are exact matches of $Z_3$ and $Z_4$, respectively, and are encoded as dictionary pointers. Chunk $Z_7$ is an approximate repeat of $Z_2$ (differing only by a substitution) but is treated as new by VLD.}
    \label{fig:source_model}
\end{figure*}

The two source models studied in this paper extend the one described in~\cite{niesen2019information} by allowing a constant number of substitution or burst-insertion edits. In both models, each source block is an edited copy of a source symbol. The output data streams $\bs_{sub}$ and $\bs_{in}$ are concatenations of source blocks under the constant-substitution and burst-insertion models, respectively. Since the source symbols are generated in the same way in both models, we describe their generation only once.

The $A$ source symbols, denoted $\gx_1,\gx_2,\ldots,\gx_A$, are i.i.d.\ binary strings generated in the following way. Fix a length distribution $\bbP_{l}$ over positive integers with mean $L$. For each $1 \le a \le A$, we draw $L_a$ from $\bbP_l$ and draw $\gx_a$ uniformly from $\Sigma^{L_a}$. The source symbols are drawn independently, and each is equally likely to be any binary string of its chosen length. The probability that $\left(\gx_1,\dotsc,\gx_A\right)=\left(\bx_1,\dotsc,\bx_A\right)$ given the lengths $L_a$ is $\prod_{a=1}^A\frac{1}{2^{L_a}}$, for any set of strings $\{\bx_a\}$ where $\bx_a$ has length $L_a$. 
Let $\cX$ denote the source symbol alphabet, i.e., $\cX=\{\gx_1,\gx_2,\ldots,\gx_A\}$. The alphabet is thus a multiset. To simplify some of the derivations, we adopt the same assumption as~\cite{niesen2019information} that $\bbP_l$ is concentrated around its mean, specifically, $\bbP_{l}\left(\frac{L}{2}\le l\le 2L\right)=1$. 
The parameters $A,B,t$, and the burst length $k$ in the insertion model, are positive integers.

In both source models, $J_1,\ldots,J_B$ are drawn independently and uniformly from $\{1,\ldots,A\}$, independently of the source alphabet. Conditional on the alphabet and these indices, the edit-position sets are selected independently across blocks, each uniformly among the $t$-element subsets of $\{1,\ldots,L_{J_b}\}$.

In the \textit{constant-substitution} source model, we flip the $t$ selected bits of $\gx_{J_b}$. The number of flipped symbols $t$ will be referred to as the \emph{substitution number}. The modified version of $\gx_{J_b}$ will be denoted $Y^{sub}_b$ and referred to as a \emph{source block}. The source string $\bs_{sub}$ is then constructed to be the concatenation of $Y^{sub}_1,\ldots,Y^{sub}_B$, i.e., $\bs_{sub}=Y^{sub}_1Y^{sub}_2\ldots Y^{sub}_B$. Note that given only $\bs_{sub}$, the boundaries between $Y^{sub}_b$ and $Y^{sub}_{b+1}$ are not known to us. The entropy of this source is denoted $H\left(\bs_{sub}\right)$, with bounds provided in the next lemma.

In the \textit{burst-insertion} source model, using the same index and position-selection rules, a string of length $k$ is inserted immediately after each selected bit of $\gx_{J_b}$. All inserted strings are drawn independently and uniformly from $\Sigma^k$, independently of the source alphabet, ancestor indices, and edit-position sets. The modified version of $\gx_{J_b}$ is denoted $Y^{in}_b$, and the source string $\bs_{in} = Y^{in}_1 Y^{in}_2 \cdots Y^{in}_B$ is defined analogously, with entropy denoted $H\left(\bs_{in}\right)$.

\begin{lem}\label{lem:entropy_sub}
Assume $L\geq 2t$. The entropy of the above source model $H\left(\bs_{sub}\right)$ satisfies
\begin{align}
    B\log\binom{\floor{L/2}}{t}\le H\left(\bs_{sub}\right)
    \le B\log \parenv{A\binom{\floor{2L}}{t}}+\left(2L+1\right)A. \label{eq:Hsub_bounds}
\end{align}
\end{lem}
\begin{IEEEproof}
Let $\bX = \left(\gx_{J_1}, \ldots, \gx_{J_B}\right)$.
For the lower bound, since $I\left(\bs_{sub}; \bX\right) \geq 0$, we have $H\left(\bs_{sub}\right) = H\left(\bs_{sub}|\bX\right) + I\left(\bs_{sub}; \bX\right) \geq H\left(\bs_{sub} | \bX\right)$. Given $\bX$, source blocks are conditionally independent and each has length $L_{J_b} \geq L/2$, so
\begin{align}
    H\left(\bs_{sub}\right) &\geq H\left(\bs_{sub} | \bX\right) = \sum_{b=1}^B H\left(Y^{sub}_b | \gx_{J_b}\right) \\
    &\geq B\log\binom{\floor{L/2}}{t}. \label{eq:Hsub_lb_pf}
\end{align}

For the upper bound, since $\bs_{sub}$ is a deterministic function of $\left(Y^{sub}_1,\ldots,Y^{sub}_B\right)$,
\begin{align}
    H\left(\bs_{sub}\right) &\leq H\left(Y^{sub}_1,\ldots,Y^{sub}_B\right) \leq H\left(Y^{sub}_1,\ldots,Y^{sub}_B|\cX\right) + H\left(\cX\right). \label{eq:Hsub_ub_pf}
\end{align}
Given $\cX$, each $Y^{sub}_b$ is determined by choosing one of $A$ source symbols and $t$ substitution positions out of at most $\floor{2L}$ positions, so $H\left(Y^{sub}_1,\ldots,Y^{sub}_B|\cX\right) \leq B\log\left(A\binom{\floor{2L}}{t}\right)$. The upper bound follows since $H\left(\cX\right)\le \left(2L+1\right)A$ by the assumption that $\bbP_l\left(L_a\le2L\right)=1$.
\end{IEEEproof}
\begin{lem}\label{lem:entropy_in}
Assume $L\geq 4t$. The entropy of the above source model $H\left(\bs_{in}\right)$ satisfies
\begin{align}
    \left(\frac{L/4-\left(t-1\right)}{L-2\left(t-1\right)}\right)^{\!t}\left(B\log\binom{\floor{L/4}}{t} + Bt\left(k-1\right)\right) \leq H\left(\bs_{in}\right) \leq B\log \parenv{A\binom{\floor{2L}}{t}} +Btk+\left(2L+1\right)A. \label{eq:Hin_bounds}
\end{align}
\end{lem}
\begin{IEEEproof}
Let $\bX = \left(\gx_{J_1}, \ldots, \gx_{J_B}\right)$. For the lower bound, we have $H\left(\bs_{in}\right) \geq H\left(\bs_{in} | \bX\right) = \sum_{b=1}^B H\left(Y^{in}_b| \gx_{J_b}\right)$. Let $\cE_b$ be the event that all $t$ selected source bits have the majority value in $\gx_{J_b}$ (choosing 0 in a tie), and every inserted burst ends with the opposite bit. Complementing the source and output bits preserves the conditional entropy, so it suffices to describe the case in which $\gx_{J_b}$ has at least as many 0s as 1s. There are then at least $L_{J_b}/2\geq L/4$ zero-valued positions, and $\cE_b$ requires that all $t$ insertions be placed immediately after a 0 and end with a 1.
Conditional on the source symbol, the last bits of the inserted bursts are independent and uniform, so the probability that all $t$ bursts end with the opposite of its majority bit is $2^{-t}$.
In the majority-zero case, the conditional probability that all selected source bits are 0 is lower bounded by $\frac{\binom{\ceil{L_{J_b}/2}}{t}}{\binom{L_{J_b}}{t}} = \prod\limits_{i=0}^{t-1}\frac{\ceil{L_{J_b}/2}-i}{L_{J_b}-i}\geq \left(\frac{L_{J_b}/2-\left(t-1\right)}{L_{J_b}-\left(t-1\right)}\right)^t \geq \left(\frac{L/4-\left(t-1\right)}{L/2-\left(t-1\right)}\right)^t$. The same bound holds for the majority-one case by bit complementation. Since the insertion locations and inserted bits are independent conditional on the source symbol, averaging this uniform bound gives $\bbP\left(\cE_b\right)\geq \left(\frac{L/4-\left(t-1\right)}{L-2\left(t-1\right)}\right)^t$. Thus,
\begin{align}
    H\left(Y^{in}_b| \gx_{J_b}\right)
    &\ge H\left(Y^{in}_b| \gx_{J_b},\cE_b\right)\bbP\left(\cE_b\right). \label{eq:Hin_lb_cond}
\end{align}
Under $\cE_b$, the insertion positions and inserted strings are uniquely recoverable by comparing the source and output from right to left. In the majority-zero case, whenever the bits disagree, the current output bit is the final 1 of a burst inserted after the current source bit, a~0. Remove and record the $k$-bit burst ending at that output position, then resume matching source bits. When the bits agree, no insertion follows the current source bit, so consume the matching pair. The majority-one case is identical after bit complementation. This reconstruction shows that distinct choices of insertion positions and remaining bits yield distinct outputs. Conditional on each fixed source symbol and $\cE_b$, the position sets among its majority-valued locations and the remaining $t\left(k-1\right)$ inserted bits are uniformly distributed. Since there are at least $\binom{\floor{L/4}}{t}$ such position sets and $2^{t\left(k-1\right)}$ choices for the remaining bits, we have $H\left(Y^{in}_b| \gx_{J_b},\cE_b\right) \geq \log\left(\binom{\floor{L/4}}{t} \cdot 2^{t\left(k-1\right)}\right)$. Summing over all $b$,
\begin{align}
    H\left(\bs_{in}\right) \geq \left(\frac{L/4-\left(t-1\right)}{L-2\left(t-1\right)}\right)^{\!t} 
    \cdot\left(B\log\binom{\floor{L/4}}{t} + Bt\left(k-1\right)\right). \label{eq:Hin_lb_sum}
\end{align}

For the upper bound, since $\bs_{in}$ is a deterministic function of $\left(Y^{in}_1,\ldots,Y^{in}_B\right)$,
\begin{align}
    H\left(\bs_{in}\right) &\leq H\left(Y^{in}_1,\ldots,Y^{in}_B\right) \leq H\left(Y^{in}_1,\ldots,Y^{in}_B|\cX\right) + H\left(\cX\right). \label{eq:Hin_ub_pf}
\end{align}
Given $\cX$, each $Y^{in}_b$ is determined by choosing one of $A$ source symbols, $t$ insertion positions out of at most $\floor{2L}$ positions, and $tk$ inserted bits, so $H\left(Y^{in}_1,\ldots,Y^{in}_B|\cX\right) \leq B\log\left(A\binom{\floor{2L}}{t}2^{tk}\right)$. The upper bound follows since $H\left(\cX\right)\le \left(2L+1\right)A$ by the assumption that $\bbP_l\left(L_a\le2L\right)=1$.
\end{IEEEproof}
For $A,L\to\infty$ and fixed positive integer $t$, the entropy bounds in Lemmas~\ref{lem:entropy_sub} and~\ref{lem:entropy_in} simplify as follows.
\begin{cor}\label{cor:entropy_asymp}
For $A,L\to\infty$ and fixed positive integer $t$,
\begin{align}\label{eq:Hsub_asymp}
H\left(\bs_{sub}\right) = Bt\log L + O\left(B\log A + AL\right),
\end{align}
\begin{align}\label{eq:Hin_asymp}
\left(4^{-t}+o\left(1\right)\right)Bt\left(k + \log L\right) \le H\left(\bs_{in}\right)
\le Bt\left(k + \log L\right) + O\left(B\log A + AL\right).
\end{align}
If also $\log A=O\left(\log L\right)$ and $AL=O\left(B\log L\right)$, then
\begin{align}
H\left(\bs_{in}\right)=\Theta_t\parenv{Bt\left(k+\log L\right)}, \label{eq:Hin_theta_t}
\end{align}
where the constants hidden by $\Theta_t$ may depend on the fixed value of $t$.
\end{cor}
\begin{IEEEproof}
For fixed $t$ and integer $N\to\infty$, $\log\binom{N}{t} = t\log N - \log\left(t!\right) + O\left(1/N\right)$. Since $\floor{L/2}=L/2+O\left(1\right)$, $\floor{L/4}=L/4+O\left(1\right)$, and $\floor{2L}=2L+O\left(1\right)$, rounding these counts does not change the asymptotic expansions below.

\emph{Substitution model:} Applying this to the lower bound in Lemma~\ref{lem:entropy_sub}, $B\log\binom{\floor{L/2}}{t} = Bt\log L + O\left(B\right)$. The upper bound gives $B\log\left(A\binom{\floor{2L}}{t}\right) + \left(2L+1\right)A = Bt\log L + O\left(B\log A + AL\right)$. Since $A\to\infty$, $O\left(B\right) \subseteq O\left(B\log A\right)$, and the result follows.

\emph{Insertion model:} The prefactor in the lower bound of Lemma~\ref{lem:entropy_in} satisfies
\begin{align}
\left(\frac{L/4-\left(t-1\right)}{L-2\left(t-1\right)}\right)^{\!t}=4^{-t}+O_t\left(L^{-1}\right),
\end{align}
while $B\log\binom{\floor{L/4}}{t}+Bt\left(k-1\right)=Bt\left(k+\log L\right)+O_t\left(B\right)$. This proves the lower bound in \eqref{eq:Hin_asymp}; the upper bound follows from Lemma~\ref{lem:entropy_in} as in the substitution case. Under the two additional assumptions, $B\log A+AL=O\left(B\log L\right)$, which is absorbed by $Bt\left(k+\log L\right)$ for fixed $t\geq1$. Combining the two sides gives \eqref{eq:Hin_theta_t}.
\end{IEEEproof}


For the subsequent asymptotic deduplication results and their appendix proofs, unless explicitly stated otherwise, we assume $B\to\infty$, $A,L\to\infty$, and a fixed positive integer $t$. For both source models, we assume $\log A=O\left(\log L\right)$ and $AL=O\left(B\log L\right)$. Furthermore, as in~\cite{lou2022data}, we consider $L=\Theta\left(B^{\alpha_1}\right)$ for a fixed $\alpha_1>0$. This polynomial relationship implies $\log B=\Theta\left(\log L\right)$ and, together with the alphabet-size assumptions, $\log A=O\left(\log B\right)$.

In the burst-insertion model, we additionally assume $k=\Omega\left(\log\left(BL\right)\right)$ and $k=o\left(L\right)$. Thus, each burst contains at least logarithmically many bits, while the number of inserted bits per block remains sublinear in the source-symbol length.

Unlike the independent-substitution model~\cite{lou2022data}, where a fixed edit probability $\delta$ gives $\delta L=\Theta\left(L\right)$ substitutions per block on average, the constant-substitution model has only $t$ substitutions per block. The alphabet-size assumptions ensure that $B\log A+AL=O\left(B\log L\right)$, so these additional terms in the entropy bounds do not change the asymptotic order. Corollary~\ref{cor:entropy_asymp} therefore gives $H\left(\bs_{sub}\right)=\Theta\left(Bt\log L\right)$ and $H\left(\bs_{in}\right)=\Theta_t\left(Bt\left(k+\log L\right)\right)$. Since $\log L=o\left(L\right)$ and $k=o\left(L\right)$, both entropies are $o\left(BL\right)$. Other asymptotic regimes, including ones in which $k$ grows more slowly, are left to future work.



Throughout, the first half consists of the first $\floor{B/2}$ source blocks, and the second half consists of the remaining blocks. We say $\gx_{J_b}$ is the ancestor of $Y^{sub}_b$ (or in the case of burst insertions, $Y^{in}_b$) and $Y^{sub}_b$ ($Y^{in}_b$) is a descendant of $\gx_{J_b}$. For each $a$, let $Y^{sub}\left(a\right)$ ($Y^{in}\left(a\right)$) denote the set $\{1 \le b \le B: J_b=a\}$ and $Y^{sub}_{\nicefrac12}\left(a\right)$ ($Y^{in}_{\nicefrac12}\left(a\right)$) denote the set $\{1\le b\le\floor{B/2}: J_b=a\}$. In other words, $Y^{sub}\left(a\right)$ ($Y^{in}\left(a\right)$) is the set of source block indexes of the descendants of $\gx_a$ and $Y^{sub}_{\nicefrac12}\left(a\right)$ ($Y^{in}_{\nicefrac12}\left(a\right)$) is the set of source block indexes of the descendants of $\gx_a$ among the first half of source blocks.

\section{Deduplication schemes}\label{sec:alg}
We describe the variable-length and multi-chunk deduplication algorithms, both studied in~\cite{niesen2019information} and restated below for completeness.

In the \emph{variable-length} deduplication algorithm, a fixed all-zero string $0^M$ of positive integer length $M$ is used as the marker to split the source string $\bs$ into chunks. Starting at the first unparsed bit, the parser closes a chunk at the end of the first occurrence of $0^M$ wholly within that chunk and restarts immediately after this marker. Any nonempty suffix containing no marker is the final chunk. Thus $\bs=Z_1\cdots Z_C$.

The encoding begins by representing the length of $\bs$ using Elias gamma coding, which encodes a positive integer $n$ in $2\floor{\log n}+1\le2\log n+1$ bits. An empty dictionary, $T_{VL}$, is initialized to store chunks during the encoding process. The chunks $\{Z_c\}_{c=1}^C$ are then processed sequentially. For each chunk, starting with $c=1$, if $Z_c$ appears for the first time, meaning $Z_c \ne Z_i$ for all $i < c$, it is encoded with an \emph{indicator bit} 1 followed by the chunk $Z_c$ itself, and it is added to the dictionary.

If $Z_c$ already exists in the dictionary, it is encoded with the indicator bit 0 followed by a pointer to the dictionary entry storing the same string. Dictionary indices start at zero, so the pointer uses $\left\lceil\log\abs{T_{VL}^{c-1}}\right\rceil\le\log\abs{T_{VL}^{c-1}}+1$ bits, where $T_{VL}^{c-1}$ is the dictionary right after processing $Z_{c-1}$. Dictionary-size logarithms are evaluated only for repeated chunks, when the dictionary is nonempty; inactive indicator terms in later encoding-cost formulas are defined as zero without evaluating these logarithms.

The total number of bits required for the variable-length deduplication algorithm to encode $\bs$ is denoted by $\cL_{VL}\left(\bs\right)$. See Fig.~\ref{fig:source_model} for an illustration of VLD chunking.

In the \emph{multi-chunk} deduplication algorithm, the source string $\bs$ is again split into chunks using the marker $0^M$, with the additional requirement that nonfinal chunks have length at least $2^{M-1}$. Starting at the first unparsed bit, the parser scans forward until the current chunk has length at least $2^{M-1}$ and ends in $0^M$, then closes the chunk and restarts. Earlier marker occurrences do not create boundaries. Any nonempty remaining suffix is the final chunk; it may be shorter than $2^{M-1}$ and need not end in a marker.
We call the chunking process \emph{multi-chunking}. We reuse $Z_1,\ldots,Z_C$ for the MCD partition, although this notation was used above for the VLD partition. 

The encoding starts with Elias gamma coding of the length of $\bs$. Chunks are encoded sequentially using an initially empty dictionary $T_{MC}$, which stores distinct chunks in order of first appearance. Consider the chunk $Z_c$.
We assume first that $Z_c$ is new, i.e., it is different from any previously appeared chunk. Let $V_c$ be the largest integer such that the chunks $Z_c,Z_{c+1},\ldots,Z_{c+V_c-1}$ are pairwise distinct and none appears among $Z_1,\ldots,Z_{c-1}$. These new chunks are bundled up and encoded as the bit 1, followed by an encoding of $V_c$ using Elias gamma coding,
followed by the binary string $Z_c Z_{c+1}\cdots Z_{c+V_c-1}$. Moreover, $Z_c,\ldots,Z_{c+V_c-1}$ are entered into the dictionary in order. The decoder identifies them by reapplying the multi-chunking rule to their concatenation, stopping after recovering $V_c$ chunks; the initial string length determines the end of any final unterminated chunk. 

On the other hand, if $Z_c$ is not new, let $\tilde c <c$ be the smallest integer satisfying $Z_{\tilde c} = Z_c$. Consider the dictionary entry containing $Z_{\tilde c}$ and the list of subsequent entries. Let $W_c$ be the largest integer such that the first $W_c$ entries in this list are equal to $Z_c, Z_{c+1},\ldots, Z_{c+W_c-1}$. Then the chunks $Z_c$ through $Z_{c+W_c-1}$ are bundled up and encoded
together as the bit 0, followed by an encoding of $W_c$ using Elias gamma coding, followed by a zero-based pointer into the dictionary entry containing chunk $Z_{\tilde c}$, using $\ceil{\log\abs{T_{MC}^{c-1}}}$ bits, where $T_{MC}^{c-1}$ is the dictionary before processing $Z_c$. After encoding either type of bundle, processing resumes at the first chunk not yet encoded.

The total number of bits for multi-chunk deduplication to encode $\bs$ is denoted $\cL_{MC}\left(\bs\right)$.
Throughout, the compression ratio is $\bbE[\abs{\bs}]/\bbE[\cL\left(\bs\right)]$, where $\cL$ denotes the compressed length under the algorithm being considered. We call an algorithm order optimal if $\bbE[\cL\left(\bs\right)]/H\left(\bs\right)=O\left(1\right)$. An optimal marker length is a positive integer $M$ minimizing the expected compressed length for the given source parameters. It is chosen before observing the realized source string.




\section{Variable-Length Deduplication}\label{sec:vld}
In this section, we analyze the VLD algorithm for the two source models.

In our asymptotic analysis, we assume $B\to \infty$ and along with it, $A,L\to \infty$ satisfying conditions described at the end of Section~\ref{sec:source}. The constants in the asymptotic notation may depend on $t$ but not on $B$, $A$, $L$, or $k$.
\subsection{Lower Bound}
\subsubsection{Constant-Substitution Model}
We start with a lower bound on the expected length of the compressed strings by variable-length deduplication under the constant substitution edit model.

\begin{restatable}{lem}{kflip}\label{lem:kflip}
Consider the constant-substitution source model of Section~\ref{sec:source} with $\log A = O\left(\log L\right)$, and variable-length deduplication with marker length $M$ satisfying $2^M\ge 2^5\log^2 L$. Let $W$ denote the total length of chunks that are stored in full at their first appearance, and set
\begin{align}
    \ell=\min\parenv{2^{M-5},L/4}.
\end{align}
For every integer $0\le \kappa\le t$, there is a constant $c_t>0$ depending only on $t$ such that
\begin{align}
    \bbE\brckt{W} \ge c_t\, L\, \ell^{\kappa} \min\parenv{A,\ \frac{B}{L^{\kappa}}}. \label{eq:kflip_W}
\end{align}
\end{restatable}
The proof of Lemma~\ref{lem:kflip} is given in Appendix~\ref{app:burst_lb}. Here $\kappa$ is an analysis parameter: each source block still has exactly $t$ substitutions, and the lemma counts a segment according to how many of those substitutions fall inside it. 

\begin{thm}\label{thm:vllbd_asymp}
Consider the constant-substitution source model of Section~\ref{sec:source}. For any integer $1\le\kappa\le t$, if $\log A = O\left(\log L\right)$ and $B=O\left(AL^\kappa\log L\right)$, then for every marker length $M$, the expected length of the compressed string by variable-length deduplication satisfies
\begin{align}
    \bbE\brckt{\cL_{VL}\left(\bs_{sub}\right)} = \Omega\parenv{\frac{BL^{\frac{1}{\kappa+1}}}{\log L}}. \label{eq:vllbd_stmt_asymp}
\end{align}
\end{thm}

\begin{IEEEproof}
Fix $1\le\kappa\le t$. Since each chunk requires at least one indicator bit and each new chunk must be stored in full, $\cL_{VL}\left(\bs_{sub}\right) \ge C + W$, where $C$ is the total number of chunks and $W$ is the total length of chunks that are stored in full at their first appearance. We show that for every $M$, at least one of $\bbE[C]$ or $\bbE[W]$ is $\Omega\left(BL^{1/\left(\kappa+1\right)}/\log L\right)$.

\textbf{Case 1: $2^M \le L^{\kappa/\left(\kappa+1\right)}$.}
Each source block $Y^{sub}_b$ has length $L_{J_b} \ge L/2$ and is an i.i.d.\ Bernoulli$\left(1/2\right)$ process. Partition each $Y^{sub}_b$ into $\floor{L_{J_b}/M} \ge \floor{L/\left(2M\right)}$ non-overlapping windows of length $M$. Each window is all zeros with probability $2^{-M}$. Every all-zero window contains at least one accepted chunk endpoint: if no chunk ends earlier within the window, the entire window lies within the current chunk and its final bit closes that chunk. Since the windows are disjoint, they provide distinct endpoints. Therefore, by linearity of expectation,
\begin{align}
    \bbE[C] \ge B\left\lfloor\frac{L}{2M}\right\rfloor \cdot \frac{1}{2^M}. \label{eq:Clbd_asymp}
\end{align}
Since $M=O\left(\log L\right)$ in this case, \eqref{eq:Clbd_asymp} gives
\begin{align}
    \bbE[C] = \Omega\parenv{\frac{BL}{M2^M}}
    = \Omega\parenv{\frac{BL^{\frac{1}{\kappa+1}}}{\log L}}. \label{eq:kappa_C_case}
\end{align}

\textbf{Case 2: $2^M > L^{\kappa/\left(\kappa+1\right)}$.}
Let $\ell=\min\left(2^{M-5},L/4\right)$. Since $\kappa\ge1$, $L^{\kappa/\left(\kappa+1\right)}$ dominates $\log^2L$, so $2^M\ge2^5\log^2L$ for all sufficiently large $L$. Lemma~\ref{lem:kflip} therefore applies. Moreover, $\ell\ge L^{\kappa/\left(\kappa+1\right)}/32$ for all sufficiently large $L$, because both $2^{M-5}$ and $L/4$ are at least this large up to constant factors. Hence,
\begin{align}
    \bbE\brckt{W}
    &\ge c_t L\ell^\kappa \min\parenv{A,\frac{B}{L^\kappa}}. \label{eq:kappa_W_lemma}
\end{align}
The assumption $B=O\left(AL^\kappa\log L\right)$ implies
\begin{align}
    \min\parenv{A,\frac{B}{L^\kappa}}
    = \Omega\parenv{\frac{B}{L^\kappa\log L}}. \label{eq:kappa_min_bound}
\end{align}
Combining \eqref{eq:kappa_W_lemma} and \eqref{eq:kappa_min_bound},
\begin{align}
    \bbE\brckt{W}
    &\ge \Omega\parenv{\frac{B L\parenv{\ell/L}^{\kappa}}{\log L}}
     = \Omega\parenv{\frac{B L^{\frac{1}{\kappa+1}}}{\log L}}. \label{eq:kappa_W_case}
\end{align}

In both cases, $\bbE[\cL_{VL}\left(\bs_{sub}\right)] = \Omega\left(BL^{1/\left(\kappa+1\right)}/\log L\right)$.
\end{IEEEproof}

\begin{cor}\label{cor:vllbd_special_cases}
Consider the constant-substitution source model of Section~\ref{sec:source} with $\log A=O\left(\log L\right)$. The following lower bounds hold for every marker length $M$:
\begin{enumerate}
    \item If $B=O\left(AL\log L\right)$, then
    \begin{align}
        \bbE\brckt{\cL_{VL}\left(\bs_{sub}\right)}
        = \Omega\parenv{\frac{B\sqrt{L}}{\log L}}. \label{eq:vllbd_kappa_one}
    \end{align}
    \item If $B=O\left(AL^t\log L\right)$, then
    \begin{align}
        \bbE\brckt{\cL_{VL}\left(\bs_{sub}\right)}
        = \Omega\parenv{\frac{BL^{\frac{1}{t+1}}}{\log L}}. \label{eq:vllbd_kappa_t}
    \end{align}
\end{enumerate}
\end{cor}
\begin{IEEEproof}
The two claims follow from Theorem~\ref{thm:vllbd_asymp} by taking $\kappa=1$ and $\kappa=t$, respectively.
\end{IEEEproof}

\subsubsection{Burst-Insertion Model}
Under the burst-insertion model, the performance of VLD depends on the chunk length, which is controlled by the marker length $M$.
We first show through Lemma~\ref{lem:allStrAppear_asymp}, Lemma~\ref{lem:numLongChunks_asymp}, and Theorem~\ref{thm:vld_lb_1_asymp} that, in the regime $k=\Omega\left(\log\left(Bt+AL\right)\right)$, where insertion lengths are at least logarithmic, small values for $M$ lead to an oversized dictionary, resulting in a \emph{compression ratio} $\frac{\bbE[|\bs_{in}|]}{\bbE[\cL_{VL}\left(\bs_{in}\right)]}$ upper bounded by a universal constant.

\begin{restatable}{lem}{allStrAppearAsymp}\label{lem:allStrAppear_asymp}
    Consider the burst-insertion source model of Section~\ref{sec:source}, and let $N,M$ be positive integers.
    Let $\cE_{\nicefrac{1}{2}}$ denote the event that at least half of the distinct strings of the form $10^M\bu10^M$, where $\bu\in R_M^{N-2M-2}$, appear among the first $\left\lfloor B/2\right\rfloor$ source blocks.
    If $2M+2 \leq N \leq \min\left(k, L/2, \log\left(Bt+AL\right)/2\right)$, then
    \begin{align}
        \Pr\left(\cE_{\nicefrac{1}{2}}\right) \geq 1 - o\left(1\right). \label{eq:vld_lb1_lemma_prob_asymp}
    \end{align}
\end{restatable}

The proof of this lemma can be found in Appendix~\ref{app:burst_lb}.

\begin{lem}[cf.~{\cite[Lemma~13]{lou2022data}}]\label{lem:numLongChunks_asymp}
    Consider variable-length chunking with positive integer marker length $M$ applied to $\bs = Y_1 Y_2 \cdots Y_B$, where $B\ge2$. Suppose that each $Y_b$, conditional on its length, is an i.i.d.\ Bernoulli$\left(1/2\right)$ string of length at least $L/2$; the source blocks need not be independent.
    If $2^M\leq L/2^8$, then, with probability at least $5/6$, at least
    \begin{align}
        \frac14\floor{\frac B2}\floor{\frac{L}{2^{M+8}}}
    \end{align}
    chunks are wholly contained in $Y_{\floor{B/2}+1} \cdots Y_B$ and have length exceeding $2^{M-4}$. In particular, their number is $\Omega\left(BL/2^M\right)$.
\end{lem}
\begin{IEEEproof}
Choose $\floor{L/2^{M+8}}$ disjoint segments of length $2^{M+7}$ in each of the last $\floor{B/2}$ source blocks. Their total number is the deterministic integer $P=\floor{B/2}\floor{L/2^{M+8}}\ge1$.

In any one segment, divide each outer quarter into two pieces of length $2^{M+4}$. The expected waiting time for $10^M$ in fair bits is $2^{M+1}$~\cite[Chapter~8]{sedgewick2013introduction}, so Markov's inequality bounds the probability that a piece contains no $10^M$ by $1/8$. The two pieces within a quarter are independent, so that quarter contains no $10^M$ with probability at most $1/64$. Both outer quarters therefore contain $10^M$ with probability at least $1-1/32$. Its leading $1$ ensures that the final $0^M$ is accepted by the VLD parser regardless of the preceding context.

For $M\le6$, every complete chunk has length at least $M>2^{M-4}$, so these two markers certify a complete chunk inside the segment. For $M\ge7$, also require that no $0^M$ ends at any of $2^{M-4}$ consecutive positions around the segment's midpoint. A union bound gives a failure probability at most $2^{M-4}2^{-M}=1/16$. On this event, a complete chunk between the outer-quarter markers bridges those positions and has length exceeding $2^{M-4}$. Thus each segment fails to certify a qualifying chunk with probability at most $1/32+1/16\le1/8$.

Let $F$ be the number of failing segments. Linearity of expectation gives $\bbE[F]\le P/8$, without requiring independence between segments. Markov's inequality gives $\Pr\left(F>3P/4\right)\le1/6$. Hence at least $P/4$ segments certify qualifying chunks with probability at least $5/6$. These chunks are distinct occurrences, since they lie wholly inside disjoint segments. Finally, $\floor{x}\ge x/2$ for $x\ge1$ yields the stated asymptotic order.
\end{IEEEproof}




\begin{thm}\label{thm:vld_lb_1_asymp}
    Consider the burst-insertion source model of Section~\ref{sec:source}, where $k = \Omega\left(\log\left(Bt+AL\right)\right)$. If
    \[
        2^M < \min\left(16k, \frac{L}{2^8},8\log\left(Bt+AL\right)\right),
    \]
    then the expected length of the compressed string by variable-length deduplication satisfies
\begin{align}
    \bbE [\cL_{VL}\left(\bs_{in}\right)] = \Omega\left(BL\right). \label{eq:vld_lb1_bound_asymp}
\end{align}
\end{thm}

\begin{IEEEproof}
If $M\le 8$, then $2^M=O\left(1\right)$. Since the theorem assumption gives $2^M<L/2^8$, Lemma~\ref{lem:numLongChunks_asymp} implies that, with probability at least $\frac{5}{6}$, the second half of the data contains $\Omega\left(BL/2^M\right)=\Omega\left(BL\right)$ chunks. Each chunk contributes at least one bit to the VLD encoding, so $\bbE[\cL_{VL}\left(\bs_{in}\right)]=\Omega\left(BL\right)$.

It remains to consider $M\ge 9$. Set $N = 2^{M-4}$, which implies $N\ge 2M+2$ and $N \leq \min\left(k, L/2, \log\left(Bt+AL\right)/2\right)$. By Lemma~\ref{lem:allStrAppear_asymp}, with probability $1 - o\left(1\right)$, at least half of the marked patterns $10^M\bu10^M$, where $\bu\in R_M^{N-2M-2}$, appear in the first $\floor{B/2}$ source blocks. Call this event $\cE_{\nicefrac{1}{2}}$. Each such pattern yields the chunk $\bu10^M$: the leading $10^M$ synchronizes the parser, and the RLL interior and following $1$ prevent an earlier boundary. Distinct interiors yield distinct chunks, so the dictionary has at least $|R_M^{N-2M-2}|/2$ entries. A union bound over possible runs of $M$ zeros gives
\begin{align}
    |R_M^{N-2M-2}|
    \ge \parenv{1-\frac{N}{2^M}}2^{N-2M-2}
    = 2^{N-O\left(M\right)}.
\end{align}
Thus, since $N=2^{M-4}$, each dictionary pointer costs at least
\begin{align*}
    \log\left(|R_M^{N-2M-2}|/2\right)=\Omega\left(N\right)
\end{align*}
bits.

By Lemma~\ref{lem:numLongChunks_asymp}, with probability at least $\frac{5}{6}$, the second half of the data contains $\Omega\left(BL/2^M\right)$ chunks of length exceeding $N$. Call this event $\cE_{ch}$. Although $\cE_{\nicefrac{1}{2}}$ and $\cE_{ch}$ concern disjoint halves of the data, they need not be independent because the two halves share the same source alphabet. By the union bound,
\begin{align}
    \Pr\left(\cE_{\nicefrac{1}{2}} \cap \cE_{ch}\right)
    \geq 1-o\left(1\right)-\frac{1}{6}
    = \frac{5}{6}-o\left(1\right)
    = \Omega\left(1\right).
\end{align}
When both hold, each of the $\Omega\left(BL/2^M\right)$ long chunks requires $\Omega\left(N\right)$ bits. Thus
\begin{align}
    \bbE[\cL_{VL}\left(\bs_{in}\right)] = \Omega\!\parenv{\frac{BL}{2^M} \cdot N} = \Omega\left(BL\right),
\end{align}
since we have $N/2^M = 2^{-4}$.
\end{IEEEproof}

Next, using Lemma~\ref{lem:substring_match_asymp} and Theorem~\ref{thm:vld_lb_3_asymp}, we show how an oversized $M$ can lead to a large number of distinct chunks, each of which needs to be encoded in full, rendering VLD ineffective with a compression ratio upper bounded by a universal constant for $\log\left(\left(4+\frac{t}{2^9}\right)BL\right) \leq k \leq \frac{L}{2^{10}}$.

\begin{restatable}{lem}{substringMatchAsymp}\label{lem:substring_match_asymp}
    Consider the burst-insertion source string $\bs_{in}$. Let $s_i$ denote its $i$th bit and $\bs_{i,n}$ its length-$n$ substring starting at $i$, where $n$ is a positive integer. Choose valid positions $i\ne j$ using only the source lengths, ancestor indices, and insertion locations, and suppose that one of the two windows lies wholly inside an insertion. Conditional on the source alphabet, ancestor indices, insertion locations, and all stream bits outside that selected window, the probability that $\bs_{i,n}=\bs_{j,n}$ is exactly $2^{-n}$.
\end{restatable}

The proof of this lemma can be found in Appendix~\ref{app:burst_lb}.

\begin{restatable}{thm}{vld_lb_3_asymp}\label{thm:vld_lb_3_asymp}
    Consider the burst-insertion source model of Section~\ref{sec:source}, where $\log\left(\left(4+\frac{t}{2^9}\right)BL\right) \leq k \leq \frac{L}{2^{10}}$. If
    $2^M \geq \frac{L}{2^8}$, then the expected length of the compressed string by variable-length deduplication satisfies 
    \begin{align}
         \bbE [\cL_{VL}\left(\bs_{in}\right)] = \Omega\left(BL\right). \label{eq:vld_lb3_bound_asymp}
     \end{align}
\end{restatable}

\begin{IEEEproof}
When $M$ is large, chunks are long. We show that a random insertion in a source block is contained, with constant probability, in a new chunk of length $\Omega\left(\min\{2^M,L\}\right)=\Omega\left(L\right)$.

Consider an insertion chosen uniformly from the $t$ length-$k$ insertions in a given source block, with the choice made independently of all bit values.
Let $\cE_{\rm flank}$ be the event that the insertion is not placed after the last source bit and that the source bits immediately before and after the insertion are both $1$. Since the insertion position is uniform and the source block has length at least $L/2$, $\Pr\left(\cE_{\rm flank}\right)=\frac14-o\left(1\right)$. On $\cE_{\rm flank}$, one side of the insertion contains at least $L/4$ source bits; choose such a side, and set $r=\min\left(2^{M-1},\floor{L/4}-1\right)$. This integer length fits beyond the flanking bit and satisfies $r\ge L/512$ for all sufficiently large $L$, since $2^M\ge L/256$. Consider the first $r$ output bits on that side beyond the flanking bit, including any other inserted bits that fall in this neighborhood. Require that the farthest bit is $1$ and that the remaining $r-1$ bits are $M$-RLL. Conditional on $\cE_{\rm flank}$ and the insertion locations, this neighborhood event has probability at least $\frac12\parenv{1-\left(r-1\right)/2^M}\ge1/4$.

Condition on the source alphabet, ancestor indices, all insertion locations, the chosen neighboring interval, and all inserted bits outside the selected insertion. Its $k$-bit string remains uniform. By the conditional form of Lemma~\ref{lem:substring_match_asymp} and the union bound, the conditional probability that it appears elsewhere in the data is at most $\left(2BL+Btk\right)/2^k\le\frac12$. Also, the conditional probability that it is not $M$-RLL is at most $k/2^M\le\frac14$. Therefore, conditional on the flank and neighborhood events, the insertion is both new and $M$-RLL with probability at least $\frac14$.




When all these events hold, the two flanking $1$ bits and the $M$-RLL insertion prevent a marker from crossing either insertion boundary. The neighboring interval contains no marker, and its outer $1$ prevents a marker from crossing its far endpoint, including when the interval is on the left. Thus the full insertion and the neighboring interval lie in a single VLD chunk. Since the inserted $k$-bit string is new, this chunk is new and is encoded explicitly. Its length is at least $r=\Omega\left(\min\{2^M,L\}\right)=\Omega\left(L\right)$. The charged intervals lie in distinct source blocks and are disjoint; if multiple successful blocks charge the same explicit chunk, these charges are disjoint subintervals of that chunk, so the sum does not double count. Each source block therefore contributes $\Omega\left(L\right)$ explicitly-encoded bits with probability $\Omega\left(1\right)$, giving $\bbE[\cL_{VL}\left(\bs_{in}\right)]=\Omega\left(BL\right)$.
\end{IEEEproof}

Finally, through Lemma~\ref{lem:dist_c} and Theorem~\ref{thm:vld_lb_2}, we derive an intermediate-marker lower bound that includes the storage forced by the inserted bits. Combined with the small- and large-marker lower bounds, this yields the optimal-marker characterization and entropy comparison in Corollary~\ref{cor:vldins_summary}.

\begin{samepage}
\begin{restatable}{lem}{distc}\label{lem:dist_c}
    Consider variable-length deduplication applied to the burst-insertion source model of Section~\ref{sec:source}, where
    \[
        4\log\left(4BL+2Btk\right) \leq 2^M \leq \frac{L}{2^8}
        \quad\text{and}\quad k \geq \log\left(4BL+2Btk\right).
    \]
    Let $\mathcal C_{\nicefrac{1}{2}}\left(\bs_{in}\right)$ denote the set of distinct chunks of the full-stream partition completed by the end of the first $\floor{B/2}$ source blocks. There exists an absolute constant $c_1>0$ such that
    \begin{align}
        \Pr\left(\abs{\mathcal C_{\nicefrac{1}{2}}\left(\bs_{in}\right)}\ge \frac{B}{512}\right) &= 1-o\left(1\right), \label{eq:dist_c_count}\\
        \bbE\brckt{\sum_{Z\in\mathcal C_{\nicefrac{1}{2}}\left(\bs_{in}\right)}\abs{Z}} &\geq c_1B\left(2^M+tk\right). \label{eq:dist_c_length}
    \end{align}
\end{restatable}
\end{samepage}
The proof of this lemma can be found in Appendix~\ref{app:burst_lb}.

\begin{restatable}{thm}{vld_lb_2}\label{thm:vld_lb_2}
	     Consider the burst-insertion source model of Section~\ref{sec:source}, where $k \geq \log\left(4BL+2Btk\right)$. If
	    $4\log\left(4BL+2Btk\right) \leq 2^M \leq \frac{L}{2^8}$, then there exist absolute constants $c_1,c_2>0$ such that the expected length of the compressed string by variable-length deduplication satisfies
     \begin{align}
         \bbE [\cL_{VL}\left(\bs_{in}\right)]
         \geq c_1B\left(2^M+tk\right)
         + c_2 \log B\left\lfloor\frac{B}{2}\right\rfloor \left(\frac{L}{2^{M+8}}-1\right). \label{eq:vld_lb2_bound}
     \end{align}
\end{restatable}

\begin{IEEEproof}
    In the full-stream VLD encoding, let $W_1$ be the cost of all chunks completed by the end of the first $\floor{B/2}$ source blocks, together with the initial length prefix, and let $W_2$ be the cost of the remaining chunks. A chunk crossing the midpoint is assigned entirely to $W_2$, so $\cL_{VL}\left(\bs_{in}\right)=W_1+W_2$.
    Since every distinct chunk in the first half is stored explicitly when it first appears, Lemma~\ref{lem:dist_c} gives
    \begin{align}
        \bbE[W_1] \geq \bbE\brckt{\sum_{Z\in\mathcal C_{\nicefrac{1}{2}}\left(\bs_{in}\right)}\abs{Z}}\ge c_1B\left(2^M+tk\right). \label{eq:vld_lb2_w1}
    \end{align}
    
    Let
    \begin{align}
        \cE_{\rm dict}=\left\{\abs{\mathcal C_{\nicefrac{1}{2}}\left(\bs_{in}\right)}\ge \frac{B}{512}\right\}.
    \end{align}
    By Lemma~\ref{lem:dist_c}, $\Pr\left(\cE_{\rm dict}\right)=1-o\left(1\right)$. Let $\cE_{\rm ch}$ denote the event that the second half contains at least
    $\frac{1}{4}\floor{\frac{B}{2}}\left(\frac{L}{2^{M+8}}-1\right)$ chunks longer than $2^{M-4}$. By Lemma~\ref{lem:numLongChunks_asymp}, $\Pr\left(\cE_{\rm ch}\right)\ge \frac56$. Hence,
    \begin{align}
        \Pr\left(\cE_{\rm dict}\cap\cE_{\rm ch}\right)\ge \frac56-o\left(1\right). \label{eq:vld_lb2_event_prob}
    \end{align}
    On $\cE_{\rm dict}\cap\cE_{\rm ch}$, each such long second-half chunk costs at least
    \begin{align}
        \min\left\{\frac{\log\left(4BL\right)}{4},\log\left(B/512\right)+1\right\}
    \end{align}
    bits: a new chunk is stored explicitly and has length greater than $2^{M-4}\ge \frac{\log\left(4BL\right)}{4}$, while a repeated chunk is represented by a pointer into a dictionary of size at least $B/512$. Since $L\ge1$, this lower bound is $\Omega\left(\log B\right)$. Therefore, for a sufficiently small absolute constant $c_2>0$,
    \begin{align}
        \bbE[W_2] \geq c_2\log B
        \left\lfloor\frac{B}{2}\right\rfloor \left(\frac{L}{2^{M+8}}-1\right). \label{eq:vld_lb2_w2}
    \end{align}

    The desired result follows from adding the lower bounds on $\bbE[W_1]$ and $\bbE[W_2]$.
    
\end{IEEEproof}

Note that Theorems~\ref{thm:vld_lb_1_asymp} and~\ref{thm:vld_lb_2} cover the lower bound for all cases of $M$ with $2^M \leq \frac{L}{2^8}$ in the regime where $k\ge\log\left(4BL+2Btk\right)$: Theorem~\ref{thm:vld_lb_1_asymp} applies below $\min\left(16k,\frac{L}{2^8},8\log\left(Bt+AL\right)\right)$, while Theorem~\ref{thm:vld_lb_2} applies from $4\log\left(4BL+2Btk\right)$ up to $\frac{L}{2^8}$. The condition $AL=O\left(B\log L\right)$ together with $A\to\infty$ and $L=\Theta\left(B^{\alpha_1}\right)$ implies $\alpha_1\le1$. Since $k=o\left(L\right)$ and $t$ is constant, $\log\left(4BL+2Btk\right)=\log\left(BL\right)+O\left(1\right)=\left(1+\alpha_1\right)\log B+O\left(1\right)$. If $\alpha_1<1$, this is strictly below $2\log B$ asymptotically; if $\alpha_1=1$, then $A\to\infty$ gives $\log\left(Bt+AL\right)=\log B+\omega\left(1\right)$. Therefore $4\log\left(4BL+2Btk\right)\le8\log\left(Bt+AL\right)$ for all sufficiently large $B$. The inequalities $4\log\left(4BL+2Btk\right)\le16k$ and $4\log\left(4BL+2Btk\right)\le L/2^8$ follow from $k\ge\log\left(4BL+2Btk\right)$ and $\log B=o\left(L\right)$, respectively, so the small and intermediate windows overlap.
\subsection{Upper Bound}
\subsubsection{Constant Substitution Model}
\begin{thm}\label{thm:vld_ub_sub}
Consider the constant-substitution source model of Section~\ref{sec:source}. Using optimal marker length $M$, the expected length of the compressed string by variable-length deduplication satisfies
\begin{align}
    \bbE\brckt{\cL_{VL}\left(\bs_{sub}\right)} \le 2AL + O\parenv{BL^{\frac{1}{2}}\log^{\frac{1}{2}}\left(BL\right)}. \label{eq:vld_ub_sub_bound}
\end{align}
\end{thm}
\begin{IEEEproof}
The variable-length deduplication partitions the source string $\bs_{sub}$ as a random number $C$ of chunks, denoted $Z_1,\ldots,Z_C$. The length of $\bs_{sub}$ can be encoded in at most $2\log\abs{\bs_{sub}} + 1$ bits by Elias gamma coding. Let $T^{c}_{VL}$ denote the dictionary right after chunk $Z_c$ is processed ($T^0_{VL}$ denotes the initial empty dictionary). We can write
\begin{align}
    \cL_{VL}\left(\bs_{sub}\right) \le  \sum_{c=1}^C\left(\mathbbm{1}_{Z_c\in T_{VL}^{c-1}} \parenv{1 + \log\abs{T_{VL}^{c-1}}+1} + \mathbbm{1}_{Z_c\notin T_{VL}^{c-1}}\parenv{1+\abs{Z_c}}\right)+2\log\abs{\bs_{sub}} +1.\label{eq:Lvlubdsimple} 
\end{align}

We next consider a partition of $\bs_{sub}$ into a random number of ``edit blocks''. We first split $\bs_{sub}$ at all the boundaries of source blocks. Each source block $Y^{sub}_b$ is further split in the following way. For each source symbol $\gx_a$ with at least one descendant, let its first descendant be $Y^{sub}_{g\left(a\right)}$, i.e., $g\left(a\right)$ is the smallest index such that $J_{g\left(a\right)}=a$. For any other descendant $Y^{sub}_b$ of $\gx_a$, we consider the mismatches between $Y^{sub}_b$ and $Y^{sub}_{g\left(a\right)}$. Suppose $Y^{sub}_b$ differs from $Y^{sub}_{g\left(a\right)}$ in positions $c_1,c_2,\ldots,c_{m}$. Since both $Y^{sub}_b$ and $Y^{sub}_{g\left(a\right)}$ are of Hamming distance $t$ from $\gx_a$, we know that $0\le  m \le 2t$. We break $Y^{sub}_b$ into $m+1$ segments at these points. Specifically, for each $1\le j \le m$, we split between the $\left(c_j-1\right)$-th symbol and the $c_j$-th symbol. The first segment is set to be empty if $c_1=1$. These segments are referred to as edit blocks. As an example, if $\gx_a = 00000000, t = 2$, and 
\begin{align}
    Y^{sub}_{g\left(a\right)} = 01000010, \quad  Y^{sub}_b = 01100000, \label{eq:vld_ub_sub_example}
\end{align}
then we have $c_1=3, c_2 = 7$, and the edit blocks are $01,1000,00$. 

Thus, conditioned on the differences between each $Y^{sub}_b$ and its corresponding ``first descendant'' source block $Y^{sub}_{g\parenv{J_b}}$, we can partition the source string $\bs_{sub} = Y^{sub}_1 Y^{sub}_2\cdots Y^{sub}_B$ into a random number $K$ of edit blocks, denoted $D_1,\ldots, D_K$ (the boundaries of source blocks are also breakpoints). Each first descendant $Y^{sub}_{g\left(a\right)}$ has no mismatch with itself and is therefore an edit block by itself. For every observed source symbol $a$, let $\iota\left(a\right)$ be the index satisfying $D_{\iota\left(a\right)}=Y^{sub}_{g\left(a\right)}$.

We define a similar notion of interior chunks and boundary chunks as in~\cite[Theorem~3]{niesen2019information} but with respect to edit blocks and the substitutions. Consider chunks whose first symbols are in edit block $D_r$. Some of them are invariant under changes to the neighboring edit blocks $D_{r-1}$ and $D_{r+1}$ and under flipping the first bit of $D_r$. In other words, replacing either neighboring edit block by an arbitrary binary string or flipping the first bit of $D_r$ does not change the existence or content of these chunks. They are referred to as ``interior'' chunks. We denote the set of indexes of interior chunks in $D_r$ by $\cC^\circ_r$. The chunks that are not interior chunks are referred to as ``boundary'' chunks. Their existence or content may depend on the neighboring edit blocks $D_{r-1}$ and $D_{r+1}$ or on the first bit of $D_r$. We denote the set of indexes of all boundary chunks that start in $D_r$ by $\partial \cC_r$. We give examples in the following of boundary chunks (indicated by underbrackets) and interior chunks (indicated by overbrackets) when marker length~$M=3$ in various cases. Vertical bars indicate the boundaries of edit blocks. Different rows are independent examples. 
\begin{align}
\cdots & 000\underbracket{10110|00}\underbracket{\vphantom{10110|00}0101000}010\cdots  \\
\cdots & \underbracket{101100|0} \underbracket{\vphantom{101100|0 }000} \underbracket{\vphantom{101100|0 }00101000} 010\cdots\\
\cdots&000\underbracket{10|11000}\overbracket{0101000}010\cdots\\
\cdots & 000 \overbracket{1011000}|\underbracket{0101000} 010\cdots\\
\cdots &000 \underbracket{101100|0} \underbracket{\vphantom{101100|0}1000} \overbracket{11000} 10\cdots \\
\cdots &000 \underbracket{1011000}| \underbracket{\vphantom{1011000}1000} \underbracket{11000} 10\cdots \label{eq:vld_ub_sub_chunks}
\end{align}

By \eqref{eq:Lvlubdsimple},
\begin{align}
    \bbE\brckt{\cL_{VL}\left(\bs_{sub}\right)} \le  \bbE\left[\sum_{r=1}^K \parenv{\sum_{\substack{c\in\cC_r^\circ\\ Z_c\notin T_{VL}^{c-1} }} \abs{Z_c}
      + \sum_{\substack{c\in\partial\cC_r\\ Z_c\notin T_{VL}^{c-1} }}\abs{Z_c}} +C +  \sum_{Z_c\in T_{VL}^{c-1} }\parenv{1 + \log\abs{T_{VL}^{c-1}}}\right]+2\log\left(2BL\right)+1.\label{eq:ELvlubd}
\end{align}
The Elias length-prefix term is retained below and absorbed by the logarithmic terms only in the final order bound.

We first consider the interior chunks that appear for the first time. Consider the edit block $D_r$ and the source block $Y^{sub}_b$ that contains $D_r$. If $Y^{sub}_b$ is not the first source block among the descendants of $\gx_{J_b}$, then $D_r$ equals the corresponding substring of $Y^{sub}_{g\parenv{J_b}}$ at the same location except possibly at its first bit. It follows from the definition of interior chunks that any interior chunk of $D_r$ must have already appeared in that substring of $Y^{sub}_{g\parenv{J_b}}$. Thus, any interior chunk of $\bs_{sub}$ that has not appeared in the dictionary is a substring of a first descendant $Y^{sub}_{g\left(a\right)}$. The total length of all first descendants is at most $2AL$. Hence,
\begin{align}     \bbE\brckt{\sum_{r=1}^K\sum_{\substack{c\in\cC_r^\circ\\ Z_c\notin T_{VL}^{c-1} }}\abs{Z_c}} \le 2AL. \label{eq:vlubdnewint}
\end{align}

Secondly, we upper bound the lengths of boundary chunks. We adopt a similar approach as~\cite{niesen2019information}. We call an occurrence of $10^M$ \textit{internal} to an edit block $D$ if it is wholly contained in $D$ and starts after the first bit of $D$. For an edit block $D$, let $\head\left(D\right)$ denote the prefix of $D$ which ends at the last zero of the first internal $10^M$ in $D$. Let $\tail\left(D\right)$ denote the suffix of $D$ which starts at the first zero of the rightmost $0^M$ in $D$. Head or tail is defined to be $D$ itself if $D$ does not contain corresponding patterns. Consider a boundary chunk $Z$ with starting position in edit block $D_r$ and ending position in edit block $D_{r+j}$, $j\ge 0$. If $j=0$, then at least one of the following two events must hold: i) the starting position of $Z$ is in $\tail\left(D_r\right)$, ii) the ending position of $Z$ is in $\head\left(D_r\right)$. The first internal $10^M$ ends at an accepted chunk endpoint and makes subsequent parsing independent of the preceding edit blocks and the first bit of $D_r$. Moreover, the rightmost $0^M$ contains at least one accepted chunk endpoint. Thus, a chunk wholly inside $D_r$ that ends after the head and starts before the tail is interior. For a similar reason, when $j>0$, the starting position of $Z$ must be in $\tail\left(D_r\right)$ and the ending position of $Z$ must be in $\head\left(D_{r+j}\right)$. Thus, the total length of boundary chunks that start in or after $D_r$ and end in or before $D_{r+j}$ is upper bounded by the total length of $\head\left(D_r\right),\ldots,\head\left(D_{r+j}\right),\tail\left(D_r\right),\ldots,\tail\left(D_{r+j}\right)$. It follows that 
\begin{align}
    \sum_{r=1}^K \sum_{c\in \partial\cC_r}\abs{Z_c} \le \sum_{r=1}^K\parenv{\abs{\head\left(D_r\right)} + \abs{\tail\left(D_r\right)}}. \label{eq:vld_ub_sub_bdry_len}
\end{align}
Condition on the source-symbol lengths, ancestor indices, and all substitution locations. Under this conditioning, $K$ and the edit-block lengths are fixed, and the bits within each nonempty $D_r$ are independent Bernoulli$\left(1/2\right)$ variables, although different edit blocks need not be independent. In an infinite sequence of such bits, the expected waiting times for $10^M$ and $0^M$ are $2^{M+1}$ and $2^{M+1}-2$, respectively~\cite[Chapter~8]{sedgewick2013introduction}. Searching for the head's defining pattern starts after the first bit, adding one bit to the first waiting-time bound. Since the head and tail are truncated at the edit-block boundaries, their conditional expected lengths are at most $2^{M+1}+1$ and $2^{M+1}-2$, respectively. Summing these conditional bounds and then averaging gives 
\begin{align}
      \bbE\brckt{\sum_{r=1}^K\sum_{\substack{c\in\partial\cC_r\\ Z_c\notin T_{VL}^{c-1} }}\abs{Z_c}} &\le  \bbE\brckt{\sum_{r=1}^K\sum_{c\in\partial\cC_r}\abs{Z_c}} \\
     &\le \bbE\brckt{\sum_{r=1}^K \parenv{2^{M+1}+1 + 2^{M+1}-2 }} \\
     &\le \left(2t+1\right)B\parenv{2^{M+2}-1}.\label{eq:vlubdnewbdry}
\end{align}

Finally, we upper bound the remaining terms in \eqref{eq:ELvlubd}. Each chunk ends at a marker accepted by the parser, except possibly for a final markerless suffix. The expected number of occurrences of $0^M$ wholly inside source blocks is at most $BL/2^M$, since a source block of length $n$ has at most $n$ candidate starting positions, each yielding $0^M$ with probability $2^{-M}$. Accepted marker intervals are disjoint, so at most $B-1$ accepted markers cross source-block boundaries. Allowing one additional chunk for the final suffix gives $\bbE[C]\le BL/2^M+B$. Therefore,
\begin{align}
    \bbE\brckt{C + \sum_{Z_c\in T_{VL}^{c-1} }\parenv{1 + \log\abs{T_{VL}^{c-1}}}}&\le \bbE[C + C\parenv{1+\log C}]\\
    &\le \bbE\brckt{C}\parenv{3+\log\left(BL\right)}\\
    &\le B\parenv{1+\frac{L}{2^M}} \parenv{3+\log\left(BL\right)}, \label{eq:vlubdpointer}
\end{align}
where the second inequality follows from $C\le 2BL$.

Thus, by plugging \eqref{eq:vlubdnewint}, \eqref{eq:vlubdnewbdry} and \eqref{eq:vlubdpointer} in \eqref{eq:ELvlubd}, we find that $\bbE\brckt{\cL_{VL}\left(\bs_{sub}\right)}$ is upper bounded by
\begin{align}
    &2AL + B\bigg(\left(2t+1\right)2^{M+2} + 3+\log\left(BL\right)-\left(2t+1\right)+ \frac{\parenv{3+\log\left(BL\right)}L}{2^M}\bigg)+2\log\left(2BL\right)+1\\
    = &2AL + B\parenv{\left(2t+1\right)2^{M+2} + \frac{\left(3+\log\left(BL\right)\right)L}{2^M}} + O\parenv{B\log\left(BL\right)}. \label{eq:vld_ub_sub_combined}
\end{align}
Writing $x=2^M$, the two marker-dependent terms are minimized over positive real $x$ at
\begin{align*}
    x=\parenv{\frac{\left(3+\log\left(BL\right)\right)L}{4\left(2t+1\right)}}^{\frac{1}{2}}.
\end{align*}
Choose $M=\ceil{\log x}$. Since $x\le2^M<2x$, rounding increases the continuous minimum by at most a factor of $5/4$. This integer choice gives the upper bound
\begin{align}
    2AL + 5B\parenv{\parenv{2t+1}\parenv{3+\log\left(BL\right)}L}^{\frac{1}{2}} +O\parenv{B\log\left(BL\right)} = 2AL + O\parenv{BL^{\frac{1}{2}}\log^{\frac{1}{2}}\left(BL\right)}. \label{eq:vld_ub_sub_min}
\end{align}
Here $\log\left(BL\right)=o\left(L\right)$ under the standing polynomial regime, so the additive logarithmic term is absorbed by $B\sqrt{L\log\left(BL\right)}$.
\end{IEEEproof}

\begin{cor}\label{cor:vldsub_summary}
Consider the constant-substitution source model of Section~\ref{sec:source} with $\log A=O\left(\log L\right)$ and $B=O\left(AL\log L\right)$. Then, for variable-length deduplication with optimal marker length,
\begin{align}
    \Omega\parenv{\frac{B\sqrt{L}}{\log L}}
    \le \bbE\brckt{\cL_{VL}\left(\bs_{sub}\right)}
    \le 2AL+O\parenv{B\sqrt{L\log\left(BL\right)}}. \label{eq:vldsub_summary_len}
\end{align}
Moreover, for constant $t$,
\begin{align}
    H\left(\bs_{sub}\right)=Bt\log L+O\left(B\log A+AL\right). \label{eq:vldsub_summary_entropy}
\end{align}
Under the standing assumption $AL=O\left(B\log L\right)$, $\bbE\brckt{\cL_{VL}\left(\bs_{sub}\right)}=B\sqrt L$ up to polylogarithmic factors,
\begin{align}
    H\left(\bs_{sub}\right)=\Theta\left(Bt\log L\right),
\end{align}
and
\begin{align}
    \frac{\bbE\brckt{\cL_{VL}\left(\bs_{sub}\right)}}{H\left(\bs_{sub}\right)}
    = \Omega\parenv{\frac{\sqrt L}{\log^2 L}}
    \cap O\parenv{\frac{\sqrt{L\log\left(BL\right)}}{\log L}}. \label{eq:vldsub_summary_entropy_ratio}
\end{align}
Since $\log\left(BL\right)=o\left(L\right)$ under the standing polynomial regime, $\bbE\brckt{\cL_{VL}\left(\bs_{sub}\right)}=o\left(BL\right)$, so VLD achieves an asymptotically arbitrarily large compression ratio while remaining polynomially larger than the entropy.
\end{cor}
\begin{IEEEproof}
The lower bound in \eqref{eq:vldsub_summary_len} follows from Corollary~\ref{cor:vllbd_special_cases}, and the upper bound follows from Theorem~\ref{thm:vld_ub_sub}. The entropy estimate \eqref{eq:vldsub_summary_entropy} follows from Corollary~\ref{cor:entropy_asymp}. Under the standing assumption $AL=O\left(B\log L\right)$, the additive $2AL$ term in \eqref{eq:vldsub_summary_len} is within the stated polylogarithmic scale, and Lemma~\ref{lem:entropy_sub} together with \eqref{eq:vldsub_summary_entropy} gives $H\left(\bs_{sub}\right)=\Theta\left(Bt\log L\right)$ for constant $t$. The entropy-ratio bounds follow by dividing \eqref{eq:vldsub_summary_len} by this entropy. Finally, $AL=O\left(B\log L\right)=o\left(BL\right)$ and $\log\left(BL\right)=o\left(L\right)$ make the upper bound in \eqref{eq:vldsub_summary_len} $o\left(BL\right)$.
\end{IEEEproof}

\subsubsection{Burst-Insertion Model}
\begin{restatable}{thm}{vldubburst}\label{thm:vld_ub_ins}
    Consider the burst-insertion source model of Section~\ref{sec:source} under the standing assumptions that $t$ is fixed and $k=\Omega\left(\log\left(BL\right)\right)\cap o\left(L\right)$. Using optimal marker length $M$, the expected length of the compressed string by variable-length deduplication satisfies
    \begin{align}
        \bbE\brckt{\cL_{VL}\left(\bs_{in}\right)} \le O\left(AL\right) + O\parenv{BL^{\frac{1}{2}}\log^{\frac{1}{2}}\left(BL\right)}+Btk. \label{eq:vld_ub_ins_min}
    \end{align}
\end{restatable}

The proof of Theorem~\ref{thm:vld_ub_ins} is very similar to that of Theorem~\ref{thm:vld_ub_sub}. The detailed proof can be found in Appendix~\ref{app:burst_ub}.

\begin{cor}\label{cor:vldins_summary}
Consider the burst-insertion source model of Section~\ref{sec:source}, where
\begin{align}
    \log\left(\left(4+\frac{t}{2^9}\right)BL\right)
    \le k\le \frac{L}{2^{10}}.
\end{align}
For variable-length deduplication with optimal marker length,
\begin{align}
    \bbE\brckt{\cL_{VL}\left(\bs_{in}\right)}
    &=\Omega\parenv{B\sqrt{L\log B}+Btk}, \label{eq:vldins_summary_lb}\\
    \bbE\brckt{\cL_{VL}\left(\bs_{in}\right)}
    &\le O\left(AL\right)+O\parenv{B\sqrt{L\log\left(BL\right)}}+Btk. \label{eq:vldins_summary_ub}
\end{align}
Under the standing regime $AL=O\left(B\log L\right)$, $L=\Theta\left(B^{\alpha_1}\right)$, and constant $t$, these bounds simplify to
\begin{align}
    \bbE\brckt{\cL_{VL}\left(\bs_{in}\right)}
    =\Theta\parenv{B\sqrt{L\log B}+Btk}. \label{eq:vldins_summary_len}
\end{align}
Moreover, Corollary~\ref{cor:entropy_asymp} gives
\begin{align}
    \frac{\bbE\brckt{\cL_{VL}\left(\bs_{in}\right)}}{H\left(\bs_{in}\right)}
    =\Theta_t\parenv{
        \frac{\sqrt{L\log B}+tk}{t\left(k+\log L\right)}}. \label{eq:vldins_summary_entropy_ratio}
\end{align}
Consequently, this ratio diverges when $k=o\left(\sqrt{L\log B}\right)$ and remains $\Theta_t\left(1\right)$ when $k=\Omega\left(\sqrt{L\log B}\right)$.
\end{cor}
\begin{IEEEproof}
The upper bound \eqref{eq:vldins_summary_ub} is Theorem~\ref{thm:vld_ub_ins}. The displayed bounds on $k$ imply
\begin{align}
    4BL+2Btk\le \left(4+\frac{t}{2^9}\right)BL,
\end{align}
and hence $k\ge\log\left(4BL+2Btk\right)$, as required by the small- and intermediate-marker coverage. For the lower bound, write $x=2^M$. If
$4\log\left(4BL+2Btk\right)\le x\le L/2^8$, Theorem~\ref{thm:vld_lb_2} gives the term $\Omega\left(Btk\right)$. If also $x\le L/2^9$, then
\begin{align}
    \frac{L}{2^8x}-1\ge \frac{L}{2^9x},
\end{align}
so its remaining terms and the arithmetic--geometric mean inequality give
\begin{align}
    \bbE\brckt{\cL_{VL}\left(\bs_{in}\right)}
    &\ge \Omega\parenv{B\left(x+\frac{L\log B}{x}+tk\right)}=\Omega\parenv{B\sqrt{L\log B}+Btk}.
\end{align}
For $L/2^9<x\le L/2^8$, the $B2^M$ term in Theorem~\ref{thm:vld_lb_2} is $\Omega\left(BL\right)$. The small-marker coverage established after Theorem~\ref{thm:vld_lb_2}, together with Theorem~\ref{thm:vld_lb_1_asymp}, gives the same $\Omega\left(BL\right)$ lower bound below the intermediate window, while Theorem~\ref{thm:vld_lb_3_asymp} gives it for $x\ge L/2^8$. Since $\sqrt{L\log B}=o\left(L\right)$ and $tk=O\left(L\right)$ under the hypotheses, $\Omega\left(BL\right)$ implies \eqref{eq:vldins_summary_lb}. Thus the lower bound holds for every marker length and hence for the optimal one.

In the standing polynomial regime, $\log\left(BL\right)=\Theta\left(\log B\right)$ and $AL=O\left(B\log L\right)=o\left(B\sqrt{L\log B}\right)$, which proves \eqref{eq:vldins_summary_len}. Finally, \eqref{eq:Hin_theta_t} gives $H\left(\bs_{in}\right)=\Theta_t\left(Bt\left(k+\log L\right)\right)$. Dividing \eqref{eq:vldins_summary_len} by this estimate proves \eqref{eq:vldins_summary_entropy_ratio} and the two stated regimes.
\end{IEEEproof}

\section{Multi-Chunk Deduplication}\label{sec:mcd}
In the following, we derive an upper bound on the expected compressed length of the multi-chunk algorithm under both edit models, using similar techniques as in~\cite[Theorem~3]{niesen2019information}.
\subsection{Constant Substitution Model}
\begin{thm}\label{thm:mcd_ub_sub}
The expected length of the compressed strings produced by multi-chunk deduplication with optimal marker length $M$ under the constant substitution model satisfies
\begin{align}
    \bbE\brckt{\cL_{MC}\left(\bs_{sub}\right)} \le O\parenv{AL} + O\parenv{B\log\left(ABL\right)}. \label{eq:mcd_ub_sub_bound}
\end{align}
\end{thm}
\begin{IEEEproof}
As in the proof of Theorem~\ref{thm:vld_ub_sub}, we consider again the
differences between each $Y^{sub}_b$ and $Y^{sub}_{g\parenv{J_b}}$ (the first descendant of the ancestor of $Y^{sub}_b$), and break $\bs_{sub}$ into edit blocks $D_1,D_2,\ldots,D_K$.

Next, we use the definitions of interior and boundary chunks from the proof of Theorem~\ref{thm:vld_ub_sub}, adapted to multi-chunking, and treat any unterminated final chunk (one not ending at an accepted marker) as a boundary chunk. Interior chunks in an edit block form a consecutive run: once their parsing is independent of the incoming context and the first bit, each subsequent marker-terminated chunk ending inside the block has the same property. We first bound the cost of these runs, then bound the boundary cost using synchronizing heads and tails. We give examples in the following about boundary chunks (indicated by underbrackets) and interior chunks (indicated by overbrackets) in multi-chunking when marker length~$M=3$ in various cases. Vertical bars indicate the boundaries of edit blocks. Different rows are independent examples.
\begin{align}
\cdots & 000\underbracket{10110|00}\underbracket{\vphantom{10110|00}0101000}010\cdots  \\
\cdots & \underbracket{101100|0} \underbracket{\vphantom{101100|0 }0000} \underbracket{\vphantom{101100|0 }0101000} 010\cdots\\
\cdots&000\underbracket{10|11000}\overbracket{0101000}010\cdots\\
\cdots & 000 \overbracket{1011000}|\underbracket{0101000} 010\cdots\\
\cdots &000 \underbracket{101100|0} \underbracket{\vphantom{101100|0}1000} \overbracket{11000} 10\cdots \label{eq:mcd_ub_sub_chunks}
\end{align}

Splitting an encoded bundle into smaller bundles cannot decrease its encoding cost. We therefore obtain an upper bound on $\cL_{MC}\left(\bs_{sub}\right)$ by splitting bundles as needed so that every boundary chunk is encoded individually.

We consider $M$ with $2^{M-1}\le L/2$; the logarithmic choice below satisfies this for all sufficiently large parameters in the standing regime. Let $\cE_{\mathrm{sep}}^{sub}$ denote the event that any two source alphabet substrings at distinct positions or in distinct source symbols, each of length $2^{M-2}$, have Hamming distance at least $2t+1$. On this event, an interior chunk can equal an earlier chunk only at the corresponding position in a descendant of the same ancestor.  Indeed, an interior chunk lies in one source block and has length at least $2^{M-1}$. Consider the two length-$2^{M-2}$ halves of its length-$2^{M-1}$ prefix. In an equal earlier occurrence, the matching prefix crosses at most one source-block boundary, since every source block has length at least $L/2\ge 2^{M-1}$. Hence at least one of its halves lies wholly inside a source block. The corresponding ancestor substrings differ in at most $2t$ bits, so separation forces the same ancestor and offset, which aligns the entire chunk.

Assume $\cE_{\mathrm{sep}}^{sub}$ holds. We first consider the interior chunks. For every edit block $D_r$ with $\cC_r^\circ\ne\varnothing$, let $c_r=\min\cC_r^\circ$ denote the index of its first interior chunk. Let $\cA_{\rm int}$ be the set of observed source symbols $a$ for which $\abs{\cC^\circ_{\iota\left(a\right)}}>0$. For every $a\in\cA_{\rm int}$, in the edit block $D_{\iota\left(a\right)} = Y^{sub}_{g\left(a\right)}$, all interior chunks are distinct and different from any other chunk that appeared previously. Thus, they can be encoded jointly using at most
\begin{align}
    &4 + 2\log\left(\abs{\cC^\circ_{\iota\left(a\right)}}\right) + \abs{Z_{c_{\iota\left(a\right)}} Z_{c_{\iota\left(a\right)}+1} \cdots Z_{c_{\iota\left(a\right)}+\abs{\cC^\circ_{\iota\left(a\right)}}-1}}\label{eq:haint}
\end{align}
bits, where $Z_{c_{\iota\left(a\right)}}$ is the first interior chunk in $D_{\iota\left(a\right)}$.

For any $D_r$ in $Y^{sub}_b$ with $J_b=a$ and $b\ne g\left(a\right)$, all its interior chunks must have appeared in the corresponding positions in $Y^{sub}_{g\left(a\right)}$. Moreover, those are the first times these chunks appeared and entered into consecutive dictionary entries. Let $\cR_{\rm rep}$ be the set of such indices $r$ for which $\abs{\cC_r^\circ}>0$. For $r\in\cR_{\rm rep}$, suppose the interior chunks in $D_r$ are $Z_{c_r},Z_{c_r+1},\ldots,Z_{c_r+\abs{\cC^\circ_r}-1}$; they can be encoded jointly using at most
\begin{align}
    5+ 2 \log\abs{\cC^\circ_r  } + \log\abs{T_{MC}^{c_r-1}}\label{eq:intMk}
\end{align}
bits, where $T_{MC}^{c_r-1}$ denotes the dictionary right after chunk $Z_{c_r-1}$ is encoded. 


For the boundary chunks, as mentioned, we assume they are encoded individually. If a boundary chunk $Z_c$ is new, then it takes at most $4 + \abs{Z_c}$ bits. If $Z_c$ has appeared before, it takes at most $5 + \log\left(\abs{T_{MC}^{c-1}}\right)$ bits. For every nonfinal chunk, if $2^{M-1}\ge\log\left(2BL\right)$,
\begin{align}
    5 + \log\left(\abs{T_{MC}^{c-1}}\right) \le 5 +\log\left(2BL\right) \le 5 + \abs{Z_c}. \label{eq:roughbd}
\end{align}

The final chunk costs at most $5+\abs{Z_c}+\log\left(2BL\right)$ bits, allowing for either a new chunk or a pointer. Together with the Elias length prefix, it requires at most $3\log\left(2BL\right)+6$ additional bits beyond the length-based chunk costs. Thus, even if the complement of $\cE_{\mathrm{sep}}^{sub}$ holds, encoding all chunks separately gives
\begin{align}
    \cL_{MC}\left(\bs_{sub}\right) &\le \sum_{c} \parenv{5 + \abs{Z_c}}+3\log\left(2BL\right)+1 \nonumber\\
    &\le 5\left(1+\frac{2BL}{2^{M-1}}\right)+2BL+3\log\left(2BL\right)+1\le 12BL, \label{eq:12bl}
\end{align}
where $1+2BL/2^{M-1}$ bounds the number of chunks, including the possibly shorter final chunk. The last inequality holds for all sufficiently large parameters, using $2^{M-1}\ge\log\left(2BL\right)$.

Thus, by combining~\eqref{eq:haint},~\eqref{eq:intMk}\added{,}~\eqref{eq:roughbd},~\eqref{eq:12bl}, we get
\begin{align}
    \cL_{MC}\left(\bs_{sub}\right) &\le
    \sum_{a\in\cA_{\rm int}}
    \Bigl(4 + 2\log\abs{\cC^\circ_{\iota\left(a\right)}} + \abs{Z_{c_{\iota\left(a\right)}}\cdots Z_{c_{\iota\left(a\right)}+\abs{\cC^\circ_{\iota\left(a\right)}}-1}}\Bigr)   + \sum_{r\in\cR_{\rm rep}} \parenv{5 + 2\log\abs{\cC_r^\circ} + \log\abs{T_{MC}^{c_r-1}}} \\
    & \quad  + \sum_{r=1}^K \sum_{c\in\partial\cC_r} \parenv{5 + \abs{Z_c}}+ \mathbbm{1}_{\overline{\cE_{\mathrm{sep}}^{sub}}} \cdot 12BL +3\log\left(2BL\right)+6\\
    &\le A\parenv{4 + 2\log\left(2L\right) + 2L} + B\left(2t+1\right)\parenv{ 5 + 2\log\left(2L\right)+\log\left(2BL\right)}\\
    & \quad + \sum_{r=1}^K \sum_{c\in\partial\cC_r} \parenv{5 + \abs{Z_c}}+ \mathbbm{1}_{\overline{\cE_{\mathrm{sep}}^{sub}}} \cdot 12BL+3\log\left(2BL\right)+6, \label{eq:mcd_ub_sub_combined}
\end{align}
where $\abs{\cA_{\rm int}}\le A$ and $\abs{\cR_{\rm rep}}\le K$. We upper bound both $\abs{\cC^\circ_{r}}$ and $\abs{Z_{c_{\iota\left(a\right)}}\cdots Z_{c_{\iota\left(a\right)}+\abs{\cC^\circ_{\iota\left(a\right)}}-1}}$ by $2L$, upper bound $K$ by $B\left(2t+1\right)$ since there are at most $\left(2t+1\right)$ edit blocks in each source block, and upper bound the dictionary size $\abs{T_{MC}^{c_r-1}}$ by $2BL$.

It remains to bound the lengths of boundary chunks. Define $\tail\left(D\right)$ as the suffix beginning $2^{M-1}-M-1$ bits before the rightmost wholly contained occurrence of $0^M$, capped at the beginning of $D$. Define $\head\left(D\right)$ as the prefix ending at the final zero of the first wholly contained occurrence of $\bu10^M$ that begins strictly after the first bit of $D$, where $\bu\in R_M^{2^{M-1}-2}$. If the relevant pattern does not appear, the corresponding head or tail is all of $D$.

The longer RLL guard in the head forces synchronization. If this pattern begins at position $p$, any earlier marker ends at or before $p+M-2$: there is no marker wholly inside $\bu$, and the following $1$ prevents one from crossing into the terminal $0^M$. The terminal marker ends at $p+2^{M-1}+M-2$, at least $2^{M-1}$ bits later, so the parser must accept it, independently of the incoming context and the first bit of $D$. Thereafter, all marker-terminated chunks ending inside $D$ are interior. At the other end, if the rightmost marker ends at position $q$, either it is accepted, or the preceding accepted boundary is at least $q-2^{M-1}+1$. In either case, any chunk continuing beyond $D$, or an unterminated final chunk when $D$ is last, starts no earlier than the tail's first position $q-2^{M-1}+2$. Consequently, every boundary-chunk portion inside $D$ is covered by its head or tail, even for chunks spanning several edit blocks. Hence,
\begin{align}
    \sum_{r=1}^K\sum_{c\in\partial\cC_r}\parenv{5+\abs{Z_c}}
    &\le \sum_{r=1}^K\Bigl(5\abs{\partial\cC_r}+\abs{\head\left(D_r\right)} +\abs{\tail\left(D_r\right)}\Bigr). \label{eq:mcd_ub_sub_bdry_len}
\end{align}

Taking expectations in \eqref{eq:mcd_ub_sub_combined} and using \eqref{eq:mcd_ub_sub_bdry_len}, $\bbE\brckt{\cL_{MC}\left(\bs_{sub}\right)}$ is at most
\begin{align}
     & \quad A\parenv{4 + 2\log\left(2L\right) + 2L}+ B\left(2t+1\right)\parenv{ 5 + 2\log\left(2L\right)+\log\left(2BL\right)}+ \bbE\left[\sum_{r=1}^K \left(5\abs{\partial\cC_r}+\abs{\head\left(D_r\right)}+\abs{\tail\left(D_r\right)}\right)\right]  \\
    & + \bbP\parenv{\overline{\cE_{\mathrm{sep}}^{sub}}} \cdot 12BL+3\log\left(2BL\right)+6. \label{eq:ELMLubd1}
\end{align}
To handle the random edit partition, condition on the source-symbol lengths, ancestor indices, and substitution locations. Then $K$ and the edit-block lengths are fixed, and the bits within each nonempty $D_r$ are i.i.d.\ Bernoulli$\left(1/2\right)$; different edit blocks need not be independent.

We adapt the renewal argument of~\cite[Appendix~F]{niesen2019information} to the longer head. After skipping the first bit of $D_r$, extend its remaining bits to an infinite fair-bit sequence and split it at each first occurrence of $0^M$, without a minimum-length requirement. The resulting subchunk lengths are i.i.d.\ with mean $2^{M+1}-2$. A subchunk with at least $2^{M-1}-1$ bits before its terminal marker contains the required head pattern: the last such bit is $1$, and the preceding $2^{M-1}-2$ bits contain no $0^M$. By a union bound over the $2^{M-1}-1$ possible marker starts, the probability of such a subchunk is at least
\begin{align}
    1-\left(2^{M-1}-1\right)2^{-M}\ge\frac12.
\end{align}
The number of subchunks through the first such occurrence therefore has mean at most $2$. Wald's identity bounds the expected head length by $1+2\left(2^{M+1}-2\right)\le2^{M+2}$. Truncating the sequence at the end of $D_r$ can only shorten the head. The reverse waiting-time argument for the tail is unchanged. Thus, under this conditioning,
\begin{align}
    \bbE\brckt{\abs{\head\left(D_r\right)} } &\le 2^{M+2},\\
    \bbE\brckt{\abs{\tail\left(D_r\right)}}&\le 2^{M+1}+2^{M-1}-M-2 \le 2^{M+2}. \label{eq:mcd_ub_sub_head_tail}
\end{align}
Moreover, every nonfinal chunk has length at least $2^{M-1}$. Counting the final chunk separately and taking unconditional expectations gives
\begin{align}
  \bbE\brckt{\sum_{r=1}^K\abs{\partial\cC_r}}
  &\le 1+\frac{1}{2^{M-1}}\bbE\Bigg[\sum_{r=1}^K\Bigl(\abs{\head\left(D_r\right)} +\abs{\tail\left(D_r\right)}\Bigr)\Bigg]. \label{eq:mcd_ub_sub_bdry_count}
\end{align}
The two conditional bounds in \eqref{eq:mcd_ub_sub_head_tail} sum to at most $13\cdot2^{M-1}$, and also to at most $2^{M+3}$. Since $K\le B\left(2t+1\right)$ deterministically, averaging over the conditioning yields
\begin{align}
    &\bbE\left[\sum_{r=1}^K\left(5\abs{\partial\cC_r}+\abs{\head\left(D_r\right)}+\abs{\tail\left(D_r\right)}\right)\right]\le 5+B\left(2t+1\right)\left(2^{M+3}+5\cdot13\right). \label{eq:douchkubd}
\end{align}
On the other hand, by the same Hamming-ball and union-bound argument used to derive \eqref{eq:kflip_p2}, 
\begin{align}
    \bbP\parenv{\overline{\cE_{\mathrm{sep}}^{sub}} } \le & \left(2AL\right)^2 \frac{\sum_{m=0}^{2t} \binom{2^{M-2}}{m}}{2^{2^{M-2}}} \\
    \le & \left(2AL\right)^2 \frac{\left(2^{M-2}\right)^{2t}}{2^{2^{M-2}}} \\
    \le & \left(2AL\right)^2 \frac{1}{2^{2^{M-3}}}, \label{eq:pecomubd}
\end{align}
for all $M\ge 3 +2\log\left(2t+1\right)$.

Thus, by plugging \eqref{eq:douchkubd} and \eqref{eq:pecomubd} into \eqref{eq:ELMLubd1} and absorbing the prefix and final-chunk overhead using $2^{M-1}\ge\log\left(2BL\right)$, we get 
\begin{align}
\bbE\brckt{\cL_{MC}\left(\bs_{sub}\right)} \le & A\parenv{6+2\log L+2L} + 48A^2BL^3 2^{-2^{M-3}} 
 + \left(2t+1\right)B\bigg(2^{M+4} + 168 
 +2\log L +\log\left(BL\right)\bigg), \label{eq:ubdmlE}
\end{align}
subject to the condition that $2^{M-1} \ge \log\left(2BL\right)$.

Pick 
\begin{align}
    M = \ceil{\log\log\parenv{\max\parenv{4A^2L^3,2BL}} +3 }. \label{eq:mcd_ub_sub_M_choice}
\end{align}
Then $2^{M-1} \ge \log\left(2BL\right)$ is satisfied. In the standing regime, $2^{M-1}=O\left(\log L\right)=o\left(L\right)$ and $M\to\infty$, so the upper constraint $2^{M-1}\le L/2$ and the condition on $M$ in \eqref{eq:pecomubd} also hold eventually. With this choice of $M$, \eqref{eq:ubdmlE} is further upper bounded by
\begin{align}
    &A\left(6+2\log L +2L\right) + 48B + \left(2t+1\right)B\left(256\log\parenv{4A^2L^3 + 2BL} + 168+2\log L+\log\left(BL\right)\right)\\
    = \ & O\parenv{AL} + O\parenv{B\log\left(ABL\right)} . \label{eq:mcd_ub_sub_final}
\end{align}
\end{IEEEproof}
By the preceding theorem and Lemma~\ref{lem:entropy_sub}, when $\log A$ and $\log B $ are both $O\left(\log L\right)$, and $AL = O\left(B\log L\right)$,
\begin{align}
    \frac{\bbE\brckt{\cL_{MC}\left(\bs_{sub}\right)}}{H\left(\bs_{sub}\right)} \le O\left(1\right). \label{eq:mcd_ub_sub_ratio}
\end{align}
Note that the $O\left(1\right)$ term depends on the substitution number $t$. Therefore, under these assumptions, MCD's expected compressed length is within a constant factor of the source entropy.

\subsection{Burst-Insertion Model}

\begin{restatable}{thm}{mcdubburst}\label{thm:mcd_ub_ins}
The expected length of the compressed strings produced by multi-chunk deduplication with optimal marker length $M$ under the burst-insertion model satisfies
\begin{align}
    \bbE\brckt{\cL_{MC}\left(\bs_{in}\right)} \le O\parenv{AL} + O\parenv{B\log\left(ABLk\right)} + O\left(Btk\right). \label{eq:mcd_ub_ins_bound}
\end{align}
\end{restatable}
The proof of Theorem~\ref{thm:mcd_ub_ins} is very similar to that of Theorem~\ref{thm:mcd_ub_sub}. The detailed proof can be found in Appendix~\ref{app:burst_ub}.
By Theorem~\ref{thm:mcd_ub_ins} and \eqref{eq:Hin_theta_t}, under the standing assumptions of Section~\ref{sec:source},
\begin{align}
    \frac{\bbE\brckt{\cL_{MC}\left(\bs_{in}\right)}}{H\left(\bs_{in}\right)}\le O_t\left(1\right), \label{eq:mcd_ub_ins_ratio}
\end{align}
where the constant may depend on the fixed insertion number $t$.

\section{Results}\label{sec:results}
In this section, we summarize the main results of the paper. 
\subsection{Variable-Length Deduplication}
Under the constant substitution source model, Theorem~\ref{thm:vllbd_asymp} gives a family of VLD lower bounds indexed by the analysis parameter $\kappa$. The choice $\kappa=t$ recovers the broad lower bound $\Omega\left(BL^{1/\left(t+1\right)}/\log L\right)$ when $B=O\left(AL^t\log L\right)$, while $\kappa=1$ gives the sharper lower bound $\Omega\left(B\sqrt L/\log L\right)$ when $B=O\left(AL\log L\right)$. Combining this latter bound with the upper bound of Theorem~\ref{thm:vld_ub_sub}, Corollary~\ref{cor:vldsub_summary} shows that VLD has expected compressed length $B\sqrt L$ up to polylogarithmic factors under the standing assumption $AL=O\left(B\log L\right)$. Under the same conditions, $\bbE[\cL_{VL}\left(\bs_{sub}\right)]=o\left(BL\right)$ while the entropy is $\Theta\left(Bt\log L\right)$, so VLD can achieve an asymptotically arbitrarily large compression ratio while remaining polynomially larger than the entropy.

Under the burst-insertion model, the performance of VLD depends on the chunk length, which is controlled by the marker length $M$. In the following summary of the burst-insertion VLD results, we assume $\log\left(\left(4+\frac{t}{2^9}\right)BL\right)\le k\le L/2^{10}$, in addition to the standing assumptions. 
The effect of $M$ on the performance of VLD is studied in Theorems~\ref{thm:vld_lb_1_asymp},~\ref{thm:vld_lb_3_asymp}, and~\ref{thm:vld_lb_2}. Theorem~\ref{thm:vld_lb_1_asymp} shows how, when $k=\Omega\left(\log\left(Bt+AL\right)\right)$ and $2^M < \min\left(16k,\frac{L}{2^8},8\log\left(Bt+AL\right)\right)$, small values of $M$ lead to an oversized dictionary, resulting in a compression ratio upper bounded by a universal constant. By Theorem~\ref{thm:vld_lb_3_asymp}, we observe that an oversized $M$ where $2^M \geq \frac{L}{2^8}$ can lead to a large number of distinct chunks, each of which needs to be encoded in full, rendering VLD ineffective with a compression ratio upper bounded by a universal constant.
Corollary~\ref{cor:vldins_summary} combines the small-, intermediate-, and large-marker regimes and shows that, for the optimal marker length, VLD is larger than the entropy by a diverging factor when $k=o\left(\sqrt{L\log B}\right)$ but is order-optimal when $k=\Omega\left(\sqrt{L\log B}\right)$.

\subsection{Multi-Chunk Deduplication}
Under the standing assumptions and the additional condition $B=O\left(AL\log L\right)$, Theorem~\ref{thm:mcd_ub_sub} shows that MCD with optimal marker length has expected compressed length within a constant factor of the source entropy. In contrast, Corollary~\ref{cor:vldsub_summary} shows that VLD's expected compressed length exceeds the entropy by a factor tending to infinity, even with its optimal marker length.
The MCD upper bound is obtained using $M=O\left(\log\log\left(ABL\right)\right)$, a smaller marker-length scale than the choice used to obtain the VLD upper bound.

Under the standing assumptions for the burst-insertion model, including $k=\Omega\left(\log\left(BL\right)\right)$ and $k=o\left(L\right)$, Theorem~\ref{thm:mcd_ub_ins} shows that MCD's expected compressed length is within a constant factor of the source entropy.
As in the constant-substitution model, the marker length used to obtain the MCD upper bound is smaller in order than that used to obtain the VLD upper bound.

\section{Conclusion}\label{sec:conclusion}
In this paper, we analyzed the performance of two deduplication algorithms, VLD and MCD, on data streams containing approximate repeats, which is a scenario frequently encountered in practice. To model such repeats, we assumed a constant number of either substitutions or burst insertions per source block.
 We showed that VLD can achieve arbitrarily large compression ratios but may also spend a number of bits larger than the entropy by an arbitrarily large factor over both proposed models. We also found that for both source models, MCD can achieve an expected compressed length within a constant factor of source entropy under mild conditions.

While this work highlights key aspects of the problem, the information-theoretic analysis of data deduplication remains rich with open questions. For instance, the source models considered here are limited to a constant number of substitutions or burst insertions, whereas real-world data often undergoes insertions, deletions, and substitutions of both single symbols and longer substrings. To develop a more comprehensive understanding, it is essential to analyze deduplication algorithms under more general source models and edit processes.

\appendices
\section{Proof of Lemmas~\ref{lem:kflip},~\ref{lem:allStrAppear_asymp},~\ref{lem:substring_match_asymp},~\ref{lem:dist_c}}\label{app:burst_lb}

\kflip*
\begin{IEEEproof}
For the segment construction, round $\ell$ down to an even integer, which affects only constant factors in the bound; we continue to write $\ell$ for this segment length. For each source symbol $\gx_a$, let $n_a$ be the number of its descendants among $Y^{sub}_1,\ldots,Y^{sub}_B$, and partition $\gx_a$ into
\begin{align}
    c_a=\floor{L_a/\ell}
\end{align}
non-overlapping length-$\ell$ segments $\bx_{a,1},\ldots,\bx_{a,c_a}$, discarding a possible shorter suffix. Since $L_a\ge L/2$ and $\ell\le L/4$, we have $c_a\ge L/\left(4\ell\right)$. Index these descendants by $i=1,\ldots,n_a$ in their order of appearance in the source stream. For each descendant of $\gx_a$, let $\bh_{a,j}^i$ denote the length-$\ell$ substring aligned with $\bx_{a,j}$, for $1\le i\le n_a$ and $1\le j\le c_a$; see Fig.~\ref{fig:bxbh}.

\begin{figure}[ht]
    \centering
    \begin{tikzpicture}[
		node distance = 0cm,
		gn/.style={minimum width=.3cm, minimum height=0.5cm, shape=rectangle,draw,font=\small,inner sep=0mm},]
	\node (y1name) [draw=none]{$\gx_a$};
	\node (y1) [gn,right = of y1name, xshift = 0.1cm, minimum width= 8.1cm]{};
	\node (ybname)[draw= none, below = of y1name, yshift = -0.4cm]{$1$};
	\node (yb)[gn, right = of ybname, xshift = 0.22cm, minimum width = 8.1 cm]{};
	\node (yBname)[draw = none, below = of ybname, yshift=-0.9cm]{$n_a$};
	\node (yB)[gn, right = of yBname, xshift = 0.1cm, minimum width = 8.1cm]{};

	\node (vdot2)[draw = none, below = of yb, yshift = -0cm]{$\vdots$};
	\node (vdot3)[draw = none, below = of ybname]{$\vdots$};
    \node (z11) [gn, minimum width = 2cm, right = of y1name,xshift = 0.1cm]{$\bx_{a,1}$};
	\node (z21) [gn, minimum width = 2cm, right = of z11]{$\bx_{a,2}$};
	\node (cdot1)[gn, minimum width = 1cm, right = of z21]{$\cdots$};
	\node (zC1)[gn, minimum width = 2cm, right = of cdot1]{$\bx_{a,c_a}$};
	\node (zcp11) [gn, draw = none,right = of zC1]{$\bx_{a,c_{a}+1}$};

	\node (z1b) [gn, minimum width = 2cm, right = of ybname,xshift = 0.22cm]{$\bh_{a,1}^1$};
	\node (z2b) [gn, minimum width = 2cm, right = of z1b]{$\bh_{a,2}^1$};
	\node (cdot2)[gn, minimum width = 1cm, right = of z2b]{$\cdots$};
	\node (zCb)[gn, minimum width = 2cm, right = of cdot2]{$\bh_{a,c_a}^1$};
	\node (zcp1b) [gn, draw=none,right = of zCb]{$\bh_{a,c_{a}+1}^1$};

	\node (z1B) [gn, minimum width = 2cm, right = of yBname,xshift = 0.1cm]{$\bh_{a,1}^{n_a}$};
	\node (z2B) [gn, minimum width = 2cm, right = of z1B]{$\bh_{a,2}^{n_a}$};
	\node (cdot3)[gn, minimum width = 1cm, right = of z2B]{$\cdots$};
	\node (zCB)[gn, minimum width = 2cm, right = of cdot3]{$\bh_{a,c_a}^{n_a}$};
	\node (zcp1B) [gn, draw=none,right = of zCB]{ $\bh_{a,c_{a}+1}^{n_a}$ };

	  \draw[decorate,decoration={brace,amplitude=5pt,raise=1pt},yshift=-1pt] ([xshift = -1cm]z11.north) -- ([xshift = 1cm]z11.north) node [black,midway, yshift = 0.5cm]
    {\footnotesize $\ell$};

	\end{tikzpicture}
    \caption{A partition of $\gx_a$ and its $n_a$ descendants into aligned segments of length $\ell$, with a possible shorter suffix. The labels $\bx_{a,c_a+1}$ and $\bh_{a,c_a+1}^i$ denote the remaining suffixes, which have length less than $\ell$ and may be empty. These suffixes are not among the $c_a$ full segments.}
    \label{fig:bxbh}
\end{figure}
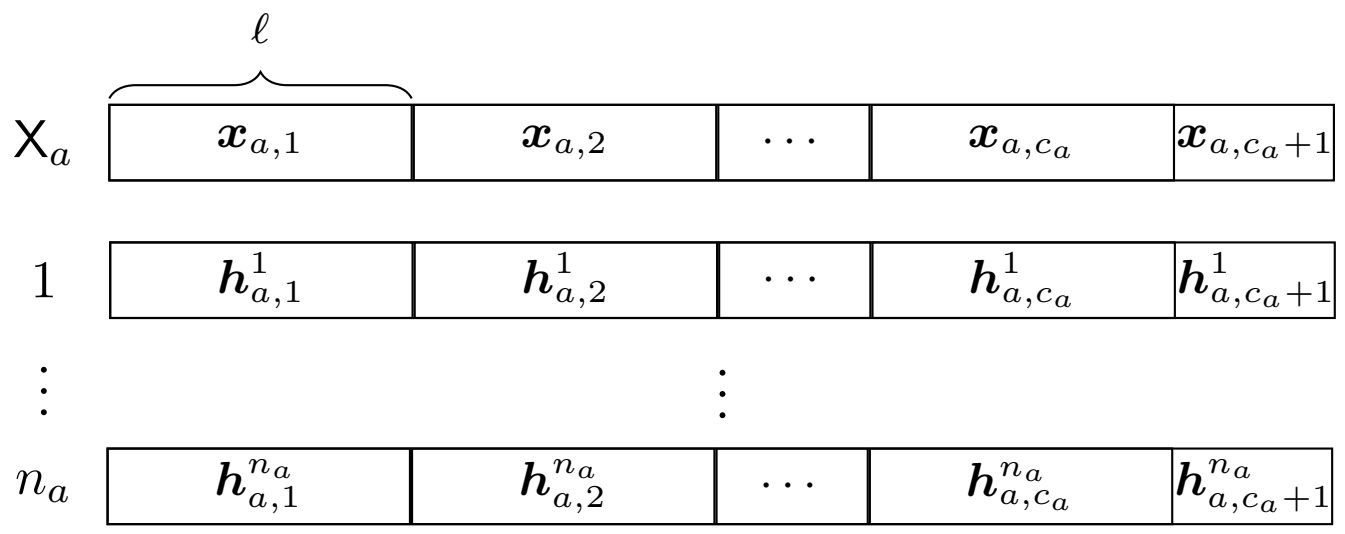

We call such a descendant segment eligible if it satisfies the following three conditions:
\begin{enumerate}
    \item $\bh_{a,j}^i$ begins and ends with $1$ and contains no run of $M$ zeros;
    \item no $\ell/2$-substring of $\bx_{a,j}$ is within Hamming distance $2t$ of an $\ell/2$-substring at a different position in the source alphabet;
    \item $\bh_{a,j}^i$ is a first appearance among its siblings, i.e., $\bh_{a,j}^i\ne \bh_{a,j}^{i'}$ for every $i'<i$.
\end{enumerate}
The first condition ensures that the entire segment lies in one VLD chunk: no marker lies inside it, and the endpoint ones prevent a marker from overlapping either endpoint.

Suppose this chunk has appeared earlier. The matching length-$\ell$ segment in that earlier chunk crosses at most one source-block boundary, since every source block has length at least $L/2\ge2\ell$. Thus at least one of its two length-$\ell/2$ halves lies wholly in a source block. The corresponding ancestral half-substrings differ in at most $2t$ positions, so the second condition forces them to have the same ancestor and the same ancestral position. This alignment places the entire matching segment at the position of $\bx_{a,j}$ in that earlier descendant, because $\bx_{a,j}$ lies wholly in its ancestor. The matching segment is therefore an earlier sibling, contradicting the third condition. Hence every eligible segment lies in a chunk stored in full at its first appearance.

The eligible segments occupy disjoint positions, so even when several lie in the same new chunk, their combined lengths do not exceed that chunk's length. Consequently, $W$ is at least $\ell$ times the number of eligible segments.

Let $D_{a,j}$ be the set of distinct strings among $\bh_{a,j}^1,\ldots,\bh_{a,j}^{n_a}$. Fix the source lengths, the indices $J_b$, and all substitution locations, and average over the source bits. Two sibling segments are equal exactly when their substitution sets within the segment are equal. Thus the fixed choices identify exactly $\abs{D_{a,j}}$ first appearances, regardless of the source bits. Each selected segment remains an i.i.d.\ Bernoulli$\left(1/2\right)$ string. Let $p_1$ and $p_2$ be uniform upper bounds on the conditional failure probabilities of the first and second conditions, respectively. Applying the union bound to each selected first appearance and then averaging over the fixed choices gives
\begin{align}
    \bbE\brckt{W}
    \ge \ell\left(1-p_1-p_2\right)\bbE\brckt{\sum_{a=1}^{A}\sum_{j=1}^{c_a}\abs{D_{a,j}}}. \label{eq:kflip_W_intermediate}
\end{align}

For the first condition, the two endpoint bits are both $1$ with probability $1/4$. A run of $M$ zeros can start at at most $\ell$ positions inside $\bh_{a,j}^i$, each with probability $2^{-M}$. Since $\ell\le2^{M-5}$, the union bound gives
\begin{align}
    p_1\le \frac34+\ell 2^{-M}\le \frac{25}{32}. \label{eq:kflip_p1}
\end{align}
For the second condition, fix an $\ell/2$-substring of $\bx_{a,j}$ and an $\ell/2$-substring at a different position in the source alphabet. Their coordinatewise differences are independent fair bits. This also holds for overlapping substrings of one ancestor: the comparison pairs at a fixed nonzero shift form disjoint paths, whose bit differences are independent. Their Hamming distance is therefore at most $2t$ with probability at most
\begin{align}
    \frac{\sum_{m=0}^{2t}\binom{\ell/2}{m}}{2^{\ell/2}}
    \le \frac{\left(\ell/2\right)^{2t}}{2^{\ell/2}},
\end{align}
for all sufficiently large $L$. There are at most $\ell$ choices inside $\bx_{a,j}$ and fewer than $2AL$ positions in the source alphabet, so
\begin{align}
    p_2\le 2AL\ell\frac{\left(\ell/2\right)^{2t}}{2^{\ell/2}}=o\left(1\right), \label{eq:kflip_p2}
\end{align}
because $2^M\ge2^5\log^2L$ gives $\ell=\Omega\left(\log^2L\right)$ even after rounding, while $\log\left(AL\right)=O\left(\log L\right)$. Together with \eqref{eq:kflip_p1}, this implies $1-p_1-p_2\ge1/8$ for all sufficiently large $L$.

It remains to lower bound the expected number of distinct segment versions. Condition on $L_1,\ldots,L_A$, so that the numbers $c_a$ of segments are fixed. Fix $\kappa$ and a set $s$ of $\kappa$ positions within segment $j$ of $\gx_a$. The conditional probability that a given source block selects $\gx_a$ and its substitution set intersects segment $j$ exactly in $s$ is
\begin{align}
    q_{\kappa} = \frac{1}{A}\cdot \frac{\binom{L_a-\ell}{t-\kappa}}{\binom{L_a}{t}} \ge \frac{c'_t}{AL^{\kappa}} \label{eq:qkappa}
\end{align}
for a constant $c'_t>0$: since $\ell\le L/4\le L_a/2$ and $L/2\le L_a\le 2L$, the numerator is at least $\parenv{L/4-t}^{t-\kappa}/\left(t-\kappa\right)!$ while the denominator is at most $\left(2L\right)^t/t!$.

When this event occurs, the corresponding descendant segment equals $\bx_{a,j}$ with exactly the bits in $s$ flipped. Distinct sets $s$ yield distinct strings, and all blocks realizing the same $s$ yield the same string. Hence $\abs{D_{a,j}}\ge \sum_s X_s$, where the sum ranges over the $\binom{\ell}{\kappa}$ subsets of size $\kappa$ and $X_s$ indicates that at least one of the $B$ blocks realizes $s$. The block choices are independent conditional on the lengths. Using $1-\left(1-q\right)^B\ge 1-e^{-Bq}\ge \left(1-e^{-1}\right)\min\left(1,Bq\right)$,
\begin{align}
    \bbE\brckt{\abs{D_{a,j}}\mid L_1,\ldots,L_A}
    &\ge \left(1-e^{-1}\right)\binom{\ell}{\kappa}\min\parenv{1, Bq_{\kappa}}. \label{eq:EDkappa}
\end{align}
The lower bound in \eqref{eq:qkappa} is uniform over the source lengths, and $\sum_a c_a\ge AL/\left(4\ell\right)$ for every realization of those lengths. Summing \eqref{eq:EDkappa} over all segments and averaging over the lengths, then substituting into \eqref{eq:kflip_W_intermediate} with $1-p_1-p_2\ge1/8$, gives
\begin{align}
    \bbE\brckt{W} &\ge \frac{1-e^{-1}}{8}\cdot\ell\cdot\frac{AL}{4\ell}\cdot\binom{\ell}{\kappa}\min\parenv{1,\frac{c'_tB}{AL^{\kappa}}}\\
    &\ge c_t\,L\,\ell^{\kappa}\min\parenv{A, \frac{B}{L^{\kappa}}}.
\end{align}
Here $\binom{\ell}{\kappa}\ge\left(\ell/t\right)^{\kappa}$ for $1\le\kappa\le t$, while $\binom{\ell}{0}=1$. The positive constant $c_t$ can be chosen uniformly over $0\le\kappa\le t$ and absorbs the factors depending only on $t$ and the initial rounding of $\ell$, so the bound holds with $\ell$ as defined in the lemma statement.
\end{IEEEproof}

\allStrAppearAsymp*
\begin{IEEEproof}
    Among the first $\left\lfloor B/2 \right\rfloor$ source blocks, partition only the $k$-bit insertions into non-overlapping pieces of length $N$. The number $P_N$ of such pieces satisfies
    \begin{align}
        P_N=\left\lfloor B/2 \right\rfloor \cdot t \cdot \left\lfloor \frac{k}{N} \right\rfloor = \Omega\left(B\right), \label{eq:allStrAppear_pieces_asymp}
    \end{align}
    since $N\le k$ and $t$ is fixed. Each piece is an i.i.d.\ Bernoulli$\left(1/2\right)$ string of length $N$. For any fixed $\bu\in R_M^{N-2M-2}$, let $\bv=10^M\bu10^M$. Then $\bv$ is an $N$-bit string, and the probability that a given piece equals $\bv$ is $2^{-N}$. Since the pieces are independent, the probability that $\bv$ does not appear in any piece is at most
    \begin{align}
        \left(1 - 2^{-N}\right)^{P_N} \le \exp\!\parenv{\!-\Omega\parenv{\frac{B}{2^N}}}. \label{eq:allStrAppear_missing_asymp}
    \end{align}
    The bounds $AL=O\left(B\log L\right)$ and $\log L=O\left(\log B\right)$, as well as the fact that $t$ is constant, are part of the standing asymptotic regime stated in Section~\ref{sec:source}. Hence the condition $N\le \log\left(Bt+AL\right)/2$ gives $2^N=O\left(\sqrt{B\log L}\right)=O\left(\sqrt{B\log B}\right)$, so $B/2^N\to\infty$. In particular, the expected number of strings $\bv=10^M \bu 10^M$ with $\bu \in R_M^{N-2M-2}$ that are missing is at most
    \begin{align}
        |R_M^{N-2M-2}| \cdot \exp\!\parenv{\!-\Omega\parenv{\frac{B}{2^N}}} = o\left(|R_M^{N-2M-2}|\right). \label{eq:allStrAppear_expfrac_asymp}
    \end{align}
    By Markov's inequality, the fraction of missing patterns exceeds $1/2$ with probability $o\left(1\right)$. Hence $\Pr\left(\cE_{\nicefrac{1}{2}}\right) \ge 1 - o\left(1\right)$.
\end{IEEEproof}

\substringMatchAsymp*

\begin{IEEEproof}
    Under the stated conditioning, the selected insertion window is uniform over $2^n$ possible strings. If the comparison window is disjoint, it is fixed, so exactly one choice matches it.

    If the windows overlap, set $d=|i-j|<n$. Equality of the windows requires bits separated by $d$ positions to agree. These equalities form chains across their union, each anchored at one of the $d$ fixed bits in the comparison window outside the selected window. They therefore determine every bit of the selected window uniquely. Exactly one of its $2^n$ possible strings satisfies all the equalities, again giving conditional probability $2^{-n}$. No independence among the outside bits is required.
\end{IEEEproof}

\AddToHookNext{env/align/begin}{%
    \renewcommand*{\theHequation}{distc-restated.\arabic{equation}}%
}
\distc*

\begin{IEEEproof}
    Set
    \begin{align}
        S=4BL+2Btk,\qquad q=\ceil{\log S},\qquad B_0=\floor{B/8}.
    \end{align}
    Since $k$ is an integer and $k\ge\log S$, we have $q\le k$. We call any $q$-bit substring lying wholly inside an insertion a \emph{core}. A core is \emph{anchored} if its first and last bits are both $1$ and it contains no run of $M$ consecutive zeros. An anchored core is fully contained in a single VLD chunk: a marker cannot lie inside it, and the endpoint ones prevent a marker from overlapping either endpoint.

    To prove the distinct-chunk count, we first find many anchored cores that have not appeared earlier. We then show that many of their source blocks contain a marker and that a later marker closes their chunks before the midpoint. Each such chunk is new, and the markers limit how many selected blocks can contribute to one chunk. For the length bound, we will charge disjoint inserted cores and nearby bits to new, completed chunks.
    
    If $U$ is an i.i.d.\ Bernoulli$\left(1/2\right)$ string of length $q$, then, for sufficiently large $B$,
    \begin{align}
        \Pr\left(U\text{ is anchored}\right)
        &\ge \Pr\left(U_1=U_q=1\right)-q2^{-\left(M+2\right)} \ge \frac{5}{32}. \label{eq:anchored_core_prob}
    \end{align}
    For each possible internal occurrence of $0^M$, the joint probability of that occurrence and both endpoint bits being $1$ is $2^{-\left(M+2\right)}$. Occurrences touching an endpoint are impossible when that endpoint is $1$. A union bound gives the first inequality, and $q/2^M\le\left(\log S+1\right)/\left(4\log S\right)$ under the hypothesis $2^M\ge4\log S$.
    
    For $1\le b\le B_0$, let $U_b$ be the first $q$ bits of the first insertion in source block $b$, ordered by its position in the stream. Let $\cF_b^-$ be the sigma-field generated by all source-symbol lengths and bits, all ancestor indices and insertion locations, and all inserted bits strictly preceding $U_b$. No bits of $U_b$ or future inserted bits are revealed. An earlier occurrence means a length-$q$ substring lying wholly before $U_b$. Conditional on $\cF_b^-$, the bits of $U_b$ remain i.i.d.\ Bernoulli$\left(1/2\right)$, and all earlier candidate substrings are determined. Moreover, fewer than $B_0\left(2L+tk\right)\le S/16$ such substrings precede $U_b$ in the stream. Hence
    \begin{align}
        \Pr\left(U_b\text{ appeared earlier}\mid \cF_b^-\right)\le \frac{1}{16}. \label{eq:core_collision_past}
    \end{align}
    Combining \eqref{eq:anchored_core_prob} and \eqref{eq:core_collision_past}, the event that $U_b$ is anchored and fresh relative to the past has conditional probability at least $3/32$.
    
    Let $G_b$ denote this event, and set $\cE_{\rm core}=\{\sum_{b=1}^{B_0}\mathbbm{1}_{G_b}>B_0/16\}$. The indicators $\mathbbm{1}_{G_b}$ need not be independent, but they satisfy
    \begin{align}
        \bbE\brckt{\mathbbm{1}_{G_b}\mid\cF_b^-}\ge \frac{3}{32}.
    \end{align}
    Every earlier $G_j$, $j<b$, is $\cF_b^-$-measurable, so the same conditional lower bound holds given the earlier indicators alone, by the tower property. Hence, the lower-tail Chernoff bound with stochastic domination~\cite[Lemma~24, Eq.~(6)]{FlinH2025}, applied with $n=B_0$, $q_b=3/32$, $\mu=3B_0/32$, and $\delta=1/3$, gives
    \begin{align}
        \Pr\parenv{\overline{\cE}_{\rm core}}
        \le \exp\parenv{-\frac{B_0}{192}}
        =e^{-\Omega\left(B\right)}. \label{eq:good_core_concentration}
    \end{align}
    We next ensure that enough of these cores lie in marker-containing blocks. Before conditioning on the realized source alphabet, fix $J_b=a$, the length $L_a$, and the insertion locations. Averaging over the source and inserted bits, $Y_b^{in}$ consists of independent fair bits and has length at least $L/2$. The expected waiting time for the first occurrence of $0^M$ in fair bits is $2^{M+1}-2$, so Markov's inequality gives
    \begin{align}
        \Pr\parenv{Y_b^{in}\text{ contains a marker}\mid J_b=a}
        \ge 1-\frac{2^{M+1}-2}{L/2}\ge\frac{63}{64}.
    \end{align}
    To justify concentration despite the shared source alphabet, condition on the source symbols. For each $a$, let
    \begin{align}
        r_a=\Pr\parenv{Y_b^{in}\text{ is markerless}\mid J_b=a,\gx_a,L_a},
    \end{align}
    where the probability is over the insertion locations and inserted bits. Averaging over the random source symbol and its length, the law of total expectation gives
    \begin{align}
        \bbE_{\gx_a,L_a}\brckt{r_a}
        &=\bbE_{\gx_a,L_a}\brckt{\Pr\parenv{Y_b^{in}\text{ markerless}\mid J_b=a,\gx_a,L_a}} =\Pr\parenv{Y_b^{in}\text{ is markerless}\mid J_b=a}
        \le \frac{1}{64}.
    \end{align}
    Let
    \begin{align}
        \bar r&=\frac{1}{A}\sum_{a=1}^A r_a, \quad  \mu_r=\frac{1}{A}\sum_{a=1}^A\bbE_{\gx_a,L_a}\brckt{r_a}\le \frac{1}{64}.\nonumber
    \end{align}
    Thus, $\bar r$ is the conditional probability that a randomly selected source block is markerless after the source alphabet is fixed. The variables $r_1,\ldots,r_A$ are independent and bounded in $[0,1]$, so the one-sided Hoeffding inequality gives
    \begin{align}
        \Pr\parenv{\bar r>\frac{1}{48}}
        &\le \Pr\parenv{\bar r-\mu_r>\frac{1}{192}} \nonumber\\
        &\le \exp\parenv{-2A\parenv{\frac{1}{192}}^2} \nonumber\\
        &=\exp\parenv{-\frac{A}{18432}}
        =e^{-\Omega\left(A\right)}=o\left(1\right). \label{eq:markerless_alphabet_concentration}
    \end{align}
    By \eqref{eq:markerless_alphabet_concentration}, the probability that the realized source alphabet has $\bar r>1/48$ is at most $e^{-A/18432}$. Fix a source alphabet with $\bar r\le 1/48$. Since the indices $J_b$ are i.i.d.\ uniform on $\{1,\ldots,A\}$ and the insertion randomness is independent across source blocks, the indicators $\mathbbm{1}\{Y_b^{in}\text{ is markerless}\}$, $1\le b\le B_0$, are conditionally i.i.d.\ Bernoulli random variables with parameter $\bar r\le 1/48$.
    Let
    \begin{align}
        Z_{\rm ml}=\sum_{b=1}^{B_0}\mathbbm{1}\{Y_b^{in}\text{ is markerless}\}.
    \end{align}
    Thus, conditional on this alphabet, $Z_{\rm ml}$ is stochastically dominated by $\operatorname{Bin}\left(B_0,1/48\right)$, whose mean is $B_0/48$. Since $B_0/32=\left(3/2\right)\left(B_0/48\right)$, Chernoff's upper-tail inequality gives
    \begin{align}
        \Pr\parenv{Z_{\rm ml}>B_0/32\mid \gx_1,L_1,\ldots,\gx_A,L_A}
        \le e^{-B_0/480}. \label{eq:markerless_block_conditional}
    \end{align}
    Let $\cE_{\rm ml}=\{Z_{\rm ml}\le B_0/32\}$. To bound its failure probability, split according to whether $\bar r>1/48$. On the favorable alphabet event, the preceding conditional Chernoff bound applies:
    \begin{align}
        \Pr\parenv{\overline{\cE}_{\rm ml}}
        &\le \Pr\parenv{\bar r>1/48}
        +\Pr\parenv{\overline{\cE}_{\rm ml},\ \bar r\le 1/48} \le e^{-A/18432}+e^{-B_0/480} =e^{-\Omega\left(A\right)}+e^{-\Omega\left(B\right)}=o\left(1\right).
        \label{eq:markerless_block_concentration}
    \end{align}
    The blocks $Y^{in}_{B_0+1},\ldots,Y^{in}_{\floor{B/2}}$ form a buffer after the selected cores. Let $\cE_{\rm close}$ be the event that at least one of these blocks contains a marker; that marker closes any chunk still open after the first $B_0$ blocks. When $\bar r\le1/48$, the conditional probability that every buffer block is markerless is at most $\left(1/48\right)^{\floor{B/2}-B_0}$. Adding the probability that $\bar r>1/48$ gives
    \begin{align}
        \Pr\parenv{\overline{\cE}_{\rm close}}\le e^{-A/18432}+\parenv{\frac{1}{48}}^{\floor{B/2}-B_0}=o\left(1\right). \label{eq:closing_marker_buffer}
    \end{align}

    We now combine the three events. On $\cE_{\rm core}\cap\cE_{\rm ml}$, more than $B_0/16-B_0/32=B_0/32$ of the first $B_0$ blocks contain both a fresh anchored core and a marker. On $\cE_{\rm close}$, each chunk containing one of these cores is completed before the second half begins. Each such chunk is new when processed: if it had appeared earlier, its core would have appeared earlier as a substring.

    A single VLD chunk can contain selected cores from at most two marker-containing blocks. If it contained cores from three, the marker in the middle block would occur between the first and third cores and terminate the chunk. Thus, on the intersection of the three events,
    \begin{align}
        \abs{\mathcal C_{\nicefrac{1}{2}}\left(\bs_{in}\right)}
        &>\frac{B_0}{64}, \nonumber\\
        \abs{\mathcal C_{\nicefrac{1}{2}}\left(\bs_{in}\right)}
        &\ge \floor{\frac{B_0}{64}}+1
        =\floor{\frac{B}{512}}+1
        >\frac{B}{512}.
    \end{align}
    The three events hold simultaneously with probability $1-o\left(1\right)$, proving \eqref{eq:dist_c_count}.
    
    It remains to bound the total length of the distinct chunks. In each argument below, an anchored fresh core certifies that its chunk is new, and $\cE_{\rm close}$ ensures that the chunk is completed before the midpoint. We charge disjoint substrings of such chunks, first from the insertions and then from neighboring bits.

    \noindent\emph{Inserted-bit contribution.} Partition every insertion in the first $B_0$ source blocks into consecutive, non-overlapping $q$-bit cores, discarding the leftover bits. Each insertion retains at least $q\floor{k/q}\ge k/2$ bits.

    For any one of these cores $U$, the preceding conditional argument gives probability at least $3/32$ of being anchored and fresh. Since $\Pr\left(\overline{\cE}_{\rm close}\right)=o\left(1\right)$, the union bound gives, for all sufficiently large $B$,
    \begin{align}
        \Pr\parenv{U\text{ anchored and fresh},\ \cE_{\rm close}}
        &\ge \frac{3}{32}-\Pr\parenv{\overline{\cE}_{\rm close}}
        \ge \frac{1}{64}.
    \end{align}
    Each qualifying core can be charged to its new, completed chunk. The cores are disjoint, so their charged lengths do not exceed the total length of these distinct chunks, even when several lie in one chunk. Linearity of expectation therefore gives
    \begin{align}
        \bbE\brckt{\sum_{Z\in\mathcal C_{\nicefrac{1}{2}}\left(\bs_{in}\right)}\abs{Z}}
        &\ge \frac{1}{64}B_0tq\floor{k/q} \ge \frac{1}{64}B_0t\frac{k}{2}
        =\Omega\left(Btk\right). \label{eq:burst_lb_btk_integrated}
    \end{align}

    \noindent\emph{Adjacent-bit contribution.} Set $m=2^{M-4}$. In each of the first $B_0$ source blocks, consider the first insertion. Since $m\le L/2^{12}$ and the underlying source symbol has length at least $L/2$, one side of this insertion contains at least $m+2$ output bits within the block for all sufficiently large $B$; choose that side deterministically from the source-block length and insertion locations. Let $\widetilde U_b$ be the $q$-bit end of the insertion nearest this side. Denote the next bit by $a_b$, the following length-$m$ substring by $I_b$, and the bit beyond $I_b$ by $g_b$. The order is
    \[
        \widetilde U_ba_bI_bg_b
        \qquad\text{or}\qquad
        g_bI_ba_b\widetilde U_b.
    \]

    Let $H_b$ be the event that $\widetilde U_b$ is anchored and fresh relative to the preceding stream, $a_b=g_b=1$, $I_b$ contains no run of $M$ zeros, and $\cE_{\rm close}$ holds. Given the insertion locations, the chosen side, and the bits $a_bI_bg_b$, the inserted core $\widetilde U_b$ remains i.i.d.\ Bernoulli$\left(1/2\right)$ and independent of the preceding stream. The same conditional argument as above therefore gives probability at least $3/32$ that it is anchored and fresh. Also,
    \begin{align}
        \Pr\parenv{a_b=g_b=1,\ I_b\text{ contains no }0^M}
        &\ge \frac14-m2^{-M}\ge\frac{3}{16}.
    \end{align}
    Requiring $\cE_{\rm close}$ costs at most $o\left(1\right)$, so
    \begin{align}
        \Pr\left(H_b\right)
        &\ge \frac{3}{16}\cdot\frac{3}{32}
        -\Pr\parenv{\overline{\cE}_{\rm close}}
        =\frac{9}{512}-o\left(1\right). \nonumber
    \end{align}

    On $H_b$, the guarded, markerless pieces $\widetilde U_ba_bI_bg_b$ (or their reverse order) lie in one new chunk completed before the midpoint. Charge the $m$ bits of $I_b$ to that chunk. The intervals $I_b$ are disjoint across source blocks, so even if several lie in the same chunk, their charged lengths do not exceed its length. Thus
    \begin{align}
        \sum_{Z\in\mathcal C_{\nicefrac{1}{2}}\left(\bs_{in}\right)}\abs{Z}
        \ge \sum_{b=1}^{B_0}m\mathbbm{1}_{H_b}.
    \end{align}
    Taking expectations gives
    \begin{align}
        \bbE\brckt{\sum_{Z\in\mathcal C_{\nicefrac{1}{2}}\left(\bs_{in}\right)}\abs{Z}}\ge \Omega\left(B_0m\right)=\Omega\left(B2^M\right). \label{eq:burst_lb_b2m_integrated}
    \end{align}
    Taking half the sum of \eqref{eq:burst_lb_btk_integrated} and \eqref{eq:burst_lb_b2m_integrated}, with a sufficiently small absolute constant, yields \eqref{eq:dist_c_length}.
\end{IEEEproof}

\section{Proof of Theorems~\ref{thm:vld_ub_ins} and~\ref{thm:mcd_ub_ins}}\label{app:burst_ub}
\vldubburst*
\begin{IEEEproof}
The variable-length deduplication partitions the source string $\bs_{in}$ as a random number $C$ of chunks, denoted $Z_1,\ldots,Z_C$. The length of $\bs_{in}$ can be encoded in at most $2\log\abs{\bs_{in}} + 1$ bits by Elias gamma coding. Let $T^{c}_{VL}$ denote the dictionary right after chunk $Z_c$ is processed ($T^0_{VL}$ denotes the initial empty dictionary). We can write
\begin{align}
    \cL_{VL}\left(\bs_{in}\right) \le&  \sum_{c=1}^C\left(\mathbbm{1}_{Z_c\in T_{VL}^{c-1}} \parenv{1 + \log\abs{T_{VL}^{c-1}}+1} + \mathbbm{1}_{Z_c\notin T_{VL}^{c-1}}\parenv{1+\abs{Z_c}}\right)+2\log\abs{\bs_{in}} +1.\label{eq:Lvlubdsimple_ins}
\end{align}

We next consider a partition of $\bs_{in}$ into a random number of ``edit blocks''. We first split $\bs_{in}$ at all the boundaries of source blocks. Each source block $Y^{in}_b$ is further split in the following way. For each source symbol $\gx_a$ with at least one descendant, let $Y^{in}_{g\left(a\right)}$ be its first descendant, i.e., $g\left(a\right)$ is the smallest index such that $J_{g\left(a\right)}=a$. All expressions involving $g\left(a\right)$ below range only over these observed source symbols, whose number is at most $A$. For any other descendant $Y^{in}_b$ of $\gx_a$, let $c_1<\cdots<c_m$ be the union of the source positions after which an insertion occurs in $Y^{in}_b$ or $Y^{in}_{g\left(a\right)}$; then $0\le m\le 2t$. Cut the source-derived bits of $Y^{in}_b$ immediately after these positions, producing at most $m+1$ edit blocks. If $Y^{in}_b$ has an insertion after $c_j$, its inserted string forms the insertion block paired with the edit block ending at $c_j$; otherwise that insertion block is empty. The insertion block paired with the final edit block is also empty. We leave each first descendant $Y^{in}_{g\left(a\right)}$ as a single edit block. As an example, for $A=1$, $B=3$, $t=2$, the edit and insertion blocks are specified in Fig.~\ref{fig:edit_blocks}.

\begin{figure}[ht]
    \centering
   \includegraphics[width=0.65\linewidth]{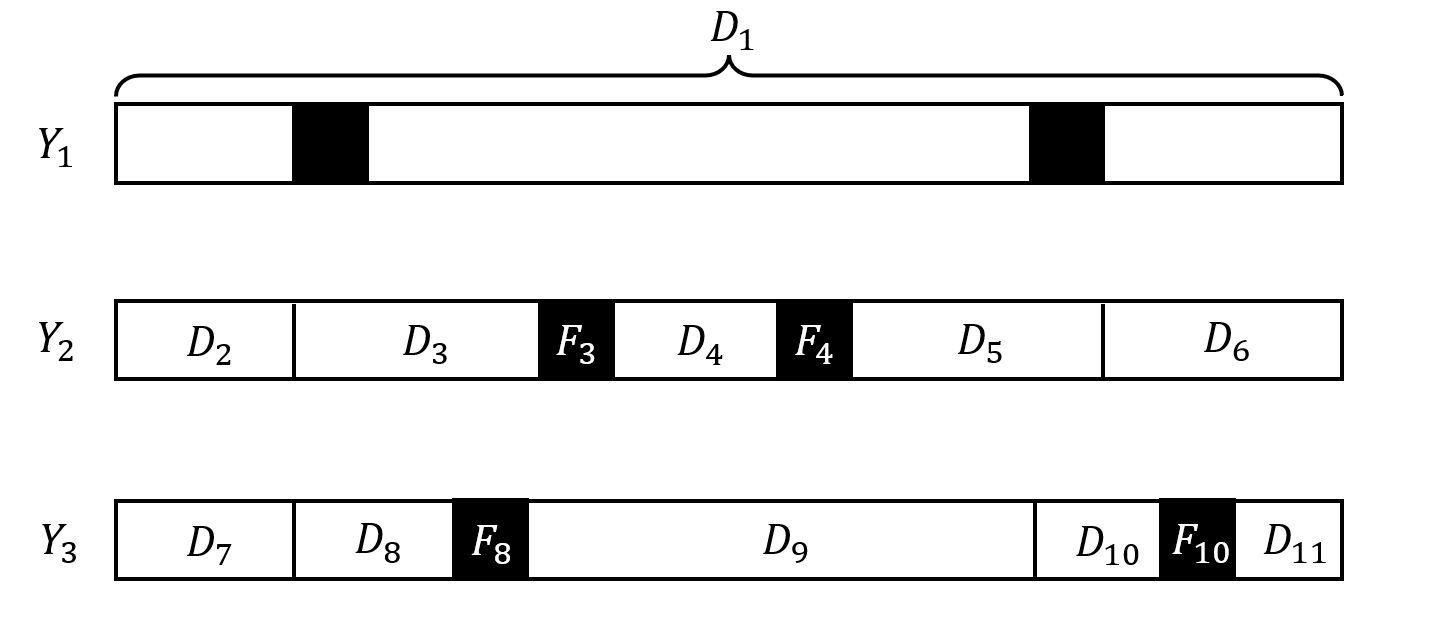}
    \caption{An example of edit and insertion blocks for $A=1$, $B=3$, $t=2$. The labels $Y_b$ abbreviate $Y_b^{in}$. The black blocks are inserted bits. Consecutive source blocks in the source string are shown in separate lines. The first descendant $Y_1$ is retained as the single edit block $D_1$, including its inserted bits. Empty insertion blocks $F_r$ are omitted from the diagram.}
    \label{fig:edit_blocks}
\end{figure}

Thus, conditioned on the differences between each $Y^{in}_b$ and its corresponding ``first descendant'' source block $Y^{in}_{g\left(J_b\right)}$, we can partition the source string $\bs_{in} = Y^{in}_1 Y^{in}_2\cdots Y^{in}_B$ into a random number $K$ of edit and insertion block pairs, denoted $D_1,F_1,\ldots, D_K,F_K$ (the boundaries of source blocks are also breakpoints). For every observed source symbol $a$, let $\iota\left(a\right)$ be the index satisfying $D_{\iota\left(a\right)}=Y^{in}_{g\left(a\right)}$.

We define a similar notion of interior chunks and boundary chunks as in~\cite[Theorem~3]{niesen2019information} but with respect to edit and insertion blocks. Consider chunks whose first symbols are in edit block $D_r$. Some of them are invariant under changes to the neighboring edit and insertion blocks. In other words, by replacing $D_{r-1}$, $F_{r-1}$, $F_r$, or $D_{r+1}$ by any other strings, the existence or content of these chunks do not change. They are referred to as ``interior'' chunks. We denote the set of indexes of interior chunks in $D_r$ by $\cC^\circ_r$. The chunks that are not interior chunks are referred to as ``boundary'' chunks. Their content depends on neighboring edit and insertion blocks $D_{r-1}$, $F_{r-1}$, $F_r$, and $D_{r+1}$. We denote the set of indexes of all boundary chunks that start in $D_r$ or $F_r$ by $\partial \cC_r$. Note that all chunks starting in an insertion block are considered boundary chunks. We give examples in the following of boundary chunks (indicated by underbrackets) and interior chunks (indicated by overbrackets) when marker length~$M=3$ in various cases. Vertical bars indicate the boundaries of edit blocks. Different rows are independent examples. 
\begin{align}
\cdots & 000\underbracket{10110|00}\underbracket{\vphantom{10110|00}0101000}010\cdots  \\
\cdots & \underbracket{101100|0} \underbracket{\vphantom{101100|0 }000} \underbracket{\vphantom{101100|0 }00101000} 010\cdots\\
\cdots&000\underbracket{10|11000}\overbracket{0101000}010\cdots\\
\cdots & 000 \overbracket{1011000}|\underbracket{0101000} 010\cdots\\
\cdots &000 \underbracket{101100|0} \underbracket{\vphantom{101100|0}1000} \overbracket{11000} 10\cdots \\
\cdots &000 \underbracket{1011000}| \underbracket{\vphantom{1011000}1000} \underbracket{11000} 10\cdots \label{eq:vld_ub_ins_chunk_examples}
\end{align}

By \eqref{eq:Lvlubdsimple_ins},
\begin{align}
    \bbE\brckt{\cL_{VL}\left(\bs_{in}\right)}
    \le \bbE\Bigg[
    \sum_{r=1}^K\sum_{\substack{c\in\cC_r^\circ\\ Z_c\notin T_{VL}^{c-1}}}\abs{Z_c} +\sum_{r=1}^K\sum_{\substack{c\in\partial\cC_r\\ Z_c\notin T_{VL}^{c-1}}}\abs{Z_c} +C+\sum_{Z_c\in T_{VL}^{c-1}}\parenv{1+\log\abs{T_{VL}^{c-1}}}\Bigg] +2\log\left(2BL+Btk\right)+1.\label{eq:ELvlubd_ins}
\end{align}
The Elias length-prefix term is $O\left(\log\left(BL+Btk\right)\right)$ and is absorbed by the final logarithmic term.

We first consider the interior chunks that appear for the first time. Consider the edit block $D_r$ and the source block $Y^{in}_b$ that contains $D_r$. If $Y^{in}_b$ is not the first source block among the descendants of $\gx_{J_b}$, then $D_r$ equals the corresponding source-derived substring of $Y^{in}_{g\left(J_b\right)}$. It follows from the definition of interior chunks that any interior chunk of $D_r$ must have already appeared in that substring of $Y^{in}_{g\left(J_b\right)}$. Thus, any interior chunk of $\bs_{in}$ that has not appeared in the dictionary is a substring of a first descendant $Y^{in}_{g\left(a\right)}$. The sum of the lengths of all first descendants is at most $A\left(2L+tk\right)$. Hence,
\begin{align}
     \bbE\brckt{\sum_{r=1}^K\sum_{\substack{c\in\cC_r^\circ\\ Z_c\notin T_{VL}^{c-1} }}\abs{Z_c}} \le A\left(2L+tk\right). \label{eq:vlubdnewint_ins}
\end{align}

Secondly, we upper bound the lengths of boundary chunks. We adopt a similar approach as~\cite{niesen2019information}. We call an occurrence of $10^M$ \textit{internal} to an edit block $D$ if it is wholly contained in $D$ and starts after the first bit of $D$. For edit block $D$, let $\head\left(D\right)$ denote the prefix of $D$ which ends at the last zero of the first internal $10^M$ in $D$. Let $\tail\left(D\right)$ denote the suffix of $D$ which starts at the first zero of the rightmost $0^M$ in $D$. Head or tail is defined to be $D$ itself if $D$ does not contain corresponding patterns. 
Consider a boundary chunk $Z$ with starting position in edit block $D_r$ and ending position in edit block $D_{r+j}$, $j\ge 0$. If $j=0$, then at least one of the following two events must hold: i) the starting position of $Z$ is in $\tail\left(D_r\right)$, ii) the ending position of $Z$ is in $\head\left(D_r\right)$. The first internal $10^M$ ends at an accepted chunk endpoint and makes subsequent parsing independent of the preceding edit and insertion blocks. Moreover, the rightmost $0^M$ contains at least one accepted chunk endpoint. Thus, a chunk wholly inside $D_r$ that ends after the head and starts before the tail is interior.
For a similar reason, when $j>0$, the starting position of $Z$ must be in $\tail\left(D_r\right)$ and the ending position of $Z$ must be in $\head\left(D_{r+j}\right)$. 
Consider a boundary chunk $Z$ with starting position in edit block $D_r$ and ending position in non-empty insertion block $F_{r+j}$ $j\ge 0$. In this case, the starting position of $Z$ is in $\tail\left(D_r\right)$.
Similarly, if the starting position of $Z$ is in the non-empty insertion block $F_r$ and the ending position is in $D_{r+j}$ $j\ge 1$, then the ending position must be in $\head\left(D_{r+j}\right)$.
Thus, boundary chunks spanning these blocks are covered by the heads and tails of the intervening edit blocks and by the intervening insertion blocks. Summing their lengths gives
\begin{align}
    \sum_{r=1}^K \sum_{c\in \partial\cC_r}\abs{Z_c} &\le \sum_{r=1}^K\Big(\abs{\head\left(D_r\right)} + \abs{\tail\left(D_r\right)} + \abs{F_r}\Big). \label{eq:vld_ub_ins_bdry_len}
\end{align}
Condition on the source-symbol lengths, ancestor indices, and all insertion locations. Under this conditioning, $K$ and the edit-block lengths are fixed, and the bits within each nonempty $D_r$ are independent Bernoulli$\left(1/2\right)$ variables, although different edit blocks need not be independent. In an infinite sequence of such bits, the expected waiting times for $10^M$ and $0^M$ are $2^{M+1}$ and $2^{M+1}-2$, respectively~\cite[Chapter~8]{sedgewick2013introduction}. Searching for the head's defining pattern starts after the first bit, adding one bit to the first waiting-time bound. Since the head and tail are truncated at the edit-block boundaries, their conditional expected lengths are at most $2^{M+1}+1$ and $2^{M+1}-2$, respectively. Summing these conditional bounds and then averaging, using $K\le B\left(2t+1\right)$ and $\sum_{r=1}^K\abs{F_r}\le Btk$, gives
\begin{align}
      \bbE\brckt{\sum_{r=1}^K\sum_{\substack{c\in\partial\cC_r\\ Z_c\notin T_{VL}^{c-1} }}\abs{Z_c}} &\le  \bbE\brckt{\sum_{r=1}^K\sum_{c\in\partial\cC_r}\abs{Z_c}} \\
     &\le \bbE\brckt{\sum_{r=1}^K \parenv{2^{M+1}+1 + 2^{M+1}-2+\abs{F_r} }} \\
     &\le \left(2t+1\right)B\parenv{2^{M+2}-1}+Btk.\label{eq:vlubdnewbdry_ins}
\end{align}

Finally, we upper bound the remaining terms in \eqref{eq:ELvlubd_ins}. Each chunk ends at a marker accepted by the parser, except possibly for a final markerless suffix. The expected number of occurrences of $0^M$ wholly inside source blocks is at most $\left(BL+Btk\right)/2^M$, since a source block of length $n$ has at most $n$ candidate starting positions, each yielding $0^M$ with probability $2^{-M}$. Accepted marker intervals are disjoint, so at most $B-1$ accepted markers cross source-block boundaries. Allowing one additional chunk for the final suffix gives $\bbE[C]\le\left(BL+Btk\right)/2^M+B$. Therefore,
\begin{align}
    \bbE\brckt{C + \sum_{Z_c\in T_{VL}^{c-1} }\parenv{1 + \log\abs{T_{VL}^{c-1}}}} \le & \bbE[C + C\parenv{1+\log C}]\\
    \le &\bbE\brckt{C}\parenv{3+\log\left(BL+Btk\right)}\\
    \le& B\parenv{1+\frac{L+tk}{2^M}} \parenv{3+\log\left(BL+Btk\right)}, \label{eq:vlubdpointer_ins}
\end{align}
where the second inequality follows from $C\le 2\left(BL+Btk\right)$.

Thus, by plugging \eqref{eq:vlubdnewint_ins}, \eqref{eq:vlubdnewbdry_ins} and \eqref{eq:vlubdpointer_ins} in \eqref{eq:ELvlubd_ins}, we find that $\bbE\brckt{\cL_{VL}\left(\bs_{in}\right)}$ is upper bounded by
\begin{align}
    &A\left(2L+tk\right) + B\bigg(\left(2t+1\right)2^{M+2} + 3+\log\left(BL+Btk\right)-\left(2t+1\right)+ \frac{\parenv{3+\log\left(BL+Btk\right)}\left(L+tk\right)}{2^M}\bigg)\nonumber\\
    &+Btk +2\log\left(2BL+Btk\right)+1\\
    =\ & A\left(2L+tk\right) + B\left(\left(2t+1\right)2^{M+2} + \frac{\left(3+\log\left(BL+Btk\right)\right)\left(L+tk\right)}{2^M}\right) + Btk + O\parenv{B\log\left(BL+Btk\right)}. \label{eq:vld_ub_ins_combined}
\end{align}
Minimizing the two $M$-dependent terms over positive real $x=2^M$ gives
\begin{align*}
    x=\parenv{\frac{\left(3+\log\left(BL+Btk\right)\right)\left(L+tk\right)}{4\left(2t+1\right)}}^{\frac{1}{2}}.
\end{align*}
Choose $M=\ceil{\log x}$. Since $x\le2^M<2x$, rounding increases the continuous minimum by at most a factor of $5/4$. Since $t$ is fixed and $k=o\left(L\right)$, the resulting upper bound is
\begin{align}
    &A\left(2L+tk\right) + 5B\left(\parenv{2t+1}\left(3+\log\left(BL+Btk\right)\right)\left(L+tk\right)\right)^{\frac{1}{2}} + Btk+O\parenv{B\log\left(BL+Btk\right)} \\
    = \ &O\left(AL\right) + O\parenv{BL^{\frac{1}{2}}\log^{\frac{1}{2}}\left(BL\right)} + Btk. \label{eq:vld_ub_ins_minimized}
\end{align}
Here $L+tk=\Theta\left(L\right)$ and $\log\left(BL+Btk\right)=\Theta\left(\log\left(BL\right)\right)$. Moreover, $k=\Omega\left(\log\left(BL\right)\right)$ and $k=o\left(L\right)$ imply $\log\left(BL\right)=o\left(L\right)$, so the $O\left(B\log\left(BL+Btk\right)\right)$ term is absorbed by $O\parenv{B\sqrt{L\log\left(BL\right)}}$.
\end{IEEEproof}

\mcdubburst*
\begin{IEEEproof}
As in the proof of Theorem~\ref{thm:vld_ub_ins}, we compare each $Y^{in}_b$ with the first descendant $Y^{in}_{g\parenv{J_b}}$ of its ancestor and use the resulting edit and insertion block pairs $D_1,F_1,\ldots,D_K,F_K$. For every observed source symbol $a$, let $\iota\left(a\right)$ satisfy $D_{\iota\left(a\right)}=Y^{in}_{g\left(a\right)}$. For brevity, set
\begin{align}
    Q=2L+tk,\qquad N=BQ=2BL+Btk, \label{eq:mcd_ub_ins_QN}
\end{align}
so every source block has length at most $Q$ and $\abs{\bs_{in}}\le N$.

We use the interior and boundary multi-chunk definitions from the proof of Theorem~\ref{thm:mcd_ub_sub}, treating every chunk that starts in an insertion block $F_r$, as well as any unterminated final chunk, as a boundary chunk. Here $\partial\cC_r$ collects the boundary chunks starting in either $D_r$ or $F_r$. Interior chunks in each $D_r$ again form a consecutive run. We bound the cost of these runs first, then the boundary cost, and finally the probability of an accidental collision.

Splitting an encoded bundle into smaller bundles cannot decrease its encoding cost. We therefore obtain an upper bound on $\cL_{MC}\left(\bs_{in}\right)$ by splitting bundles as needed so that every boundary chunk is encoded individually. 

Let $\cE_{\mathrm{col}}^{in}$ denote the event that there is at least one interior chunk $Z_c$ in a first descendant $Y^{in}_{g\left(a\right)}$ of an observed source symbol $a$ (where $Y^{in}_{g\left(a\right)}=D_{\iota\left(a\right)}$ and $c \in \cC^\circ_{\iota\left(a\right)}$) that is either equal to another interior chunk of a first descendant or equal to a boundary chunk that appears before $Z_c$.

Assume the complement of $\cE_{\mathrm{col}}^{in}$ holds. Every interior chunk of a nonfirst descendant also occurs as an interior chunk at the corresponding position of its first descendant: the edit block is an unchanged substring there, and the chunk's parsing is invariant to its neighboring bits. Thus the absence of collisions between first-descendant interior chunks also excludes collisions with earlier interior chunks of nonfirst descendants. We first consider the interior chunks. For every edit block $D_r$ with $\cC_r^\circ\ne\varnothing$, let $c_r=\min\cC_r^\circ$ denote the index of its first interior chunk. Let
\begin{align}
    \cA_{\rm int}=\left\{a:g\left(a\right)\text{ is defined and }\abs{\cC^\circ_{\iota\left(a\right)}}>0\right\}.
\end{align}
For every $a\in\cA_{\rm int}$, all interior chunks in $D_{\iota\left(a\right)} = Y^{in}_{g\left(a\right)}$ are distinct and different from every chunk that appeared previously. Thus, they can be encoded jointly using at most
\begin{align}
    4 + 2\log\left(\abs{\cC^\circ_{\iota\left(a\right)}}\right) + \abs{Z_{c_{\iota\left(a\right)}} Z_{c_{\iota\left(a\right)}+1} \cdots Z_{c_{\iota\left(a\right)}+\abs{\cC^\circ_{\iota\left(a\right)}}-1}}\label{eq:haint_ins}
\end{align}
bits, where $Z_{c_{\iota\left(a\right)}}$ is the first interior chunk in $D_{\iota\left(a\right)}$.

For any $D_r$ that lies in some $Y^{in}_b$ with $J_b=a$ and $b\ne g\parenv{J_b}$, all of its interior chunks must have appeared in the corresponding positions in $Y^{in}_{g\left(a\right)}$. Moreover, those corresponding occurrences are the first times these chunks appeared and entered consecutive dictionary entries. Let $\cR_{\rm rep}$ be the set of such indices $r$ for which $\abs{\cC_r^\circ}>0$. For $r\in\cR_{\rm rep}$, suppose the interior chunks in $D_r$ are $Z_{c_r},Z_{c_r+1},\ldots,Z_{c_r+\abs{\cC^\circ_r}-1}$. They can be encoded jointly using at most
\begin{align}
    5+ 2 \log\abs{\cC^\circ_r  } + \log\abs{T_{MC}^{c_r-1}}\label{eq:intMk_ins}
\end{align}
bits, where $T_{MC}^{c_r-1}$ denotes the dictionary right after chunk $Z_{c_r-1}$ is encoded. 


For the boundary chunks, as mentioned, we assume they are encoded individually. If a boundary chunk $Z_c$ is new, then it takes at most $4 + \abs{Z_c}$ bits. If $Z_c$ has appeared before, it takes at most $5 + \log\left(\abs{T_{MC}^{c-1}}\right)$ bits. For every nonfinal chunk, if $2^{M-1}\ge\log N$,
\begin{align}
    5 + \log\left(\abs{T_{MC}^{c-1}}\right) \le 5 +\log N \le 5 + \abs{Z_c}. \label{eq:roughbd_ins}
\end{align}

The possibly shorter final chunk can instead be encoded separately using at most $5+\abs{Z_c}+\log N$ bits, allowing for either a new chunk or a pointer. If $\cE_{\mathrm{col}}^{in}$ holds, encoding all chunks separately therefore gives
\begin{align}
    \cL_{MC}\left(\bs_{in}\right) &\le \sum_{c} \parenv{5 + \abs{Z_c}}+3\log N+1 \\
    &\le  5\left(1+\frac{N}{2^{M-1}}\right) + N+3\log N+1 \\
    &\le 6N+6+3\log N, \label{eq:12bl_ins}
\end{align}
where $1+N/2^{M-1}$ bounds the number of chunks in $\bs_{in}$ from above, including a possibly shorter final chunk. Together, the separately treated final chunk and the Elias length prefix contribute at most $3\log N+6$ additional bits to the ordinary-event bound below.

Thus, by combining~\eqref{eq:haint_ins},~\eqref{eq:intMk_ins},~\eqref{eq:roughbd_ins},~\eqref{eq:12bl_ins}, we get 
\begin{align}
    \cL_{MC}\left(\bs_{in}\right) &\le
    \sum_{a\in\cA_{\rm int}}
    \Bigl(4 + 2\log\abs{\cC^\circ_{\iota\left(a\right)}} + \abs{Z_{c_{\iota\left(a\right)}}\cdots Z_{c_{\iota\left(a\right)}+\abs{\cC^\circ_{\iota\left(a\right)}}-1}}\Bigr) \nonumber + \sum_{r\in\cR_{\rm rep}} \parenv{5 + 2\log\abs{\cC_r^\circ} + \log\abs{T_{MC}^{c_r-1}}} \\
    & \quad  + \sum_{r=1}^K \sum_{c\in\partial\cC_r} \parenv{5 + \abs{Z_c}} + \mathbbm{1}_{\cE_{\mathrm{col}}^{in}} \cdot \parenv{6N+6+3\log N} +3\log N+6\\
    &\le A\parenv{4 + 2\log Q + Q}   + B\left(2t+1\right)\parenv{ 5 + 2\log Q+\log N} + \sum_{r=1}^K \sum_{c\in\partial\cC_r} \parenv{5 + \abs{Z_c}}\\
    & \quad + \mathbbm{1}_{\cE_{\mathrm{col}}^{in}} \cdot \parenv{6N+6+3\log N} +3\log N+6, \label{eq:mcd_ub_ins_combined}
\end{align}
where $\abs{\cA_{\rm int}}\le A$, $\abs{\cR_{\rm rep}}\le K$, we upper bound both $\abs{\cC^\circ_{r}}$ and $\abs{Z_{c_{\iota\left(a\right)}}\cdots Z_{c_{\iota\left(a\right)}+\abs{\cC^\circ_{\iota\left(a\right)}}-1}}$ by $Q$, upper bound $K$ by $B\left(2t+1\right)$ since there are at most $\left(2t+1\right)$ edit blocks in each source block, and upper bound the dictionary size $\abs{T_{MC}^{c_r-1}}$ by $N$.

It remains to bound the lengths of boundary chunks. Use the head and tail defined in the proof of Theorem~\ref{thm:mcd_ub_sub}: the head ends at the first wholly contained $\bu10^M$ beginning strictly after the first bit of $D$, with $\bu\in R_M^{2^{M-1}-2}$, and the tail extends $2^{M-1}-M-1$ bits before the rightmost wholly contained $0^M$, capped at the beginning of $D$. If the relevant pattern is absent, the corresponding head or tail is all of $D$. The synchronization and coverage argument proved there applies within each edit block unchanged. Every remaining boundary-chunk portion lies in an insertion block $F_r$, so the total length of boundary chunks is bounded by the total length of the heads, tails, and insertion blocks:
\begin{align}
    \sum_{r=1}^K\sum_{c\in\partial\cC_r}\parenv{5+\abs{Z_c}}
    &\le \sum_{r=1}^K\Bigl(5\abs{\partial\cC_r}+\abs{\head\left(D_r\right)} \nonumber\\
    &\qquad+\abs{\tail\left(D_r\right)}+\abs{F_r}\Bigr).
    \label{eq:mcd_ub_ins_bdry_headtail}
\end{align}

Taking expectations in \eqref{eq:mcd_ub_ins_combined} and using \eqref{eq:mcd_ub_ins_bdry_headtail}, $\bbE\brckt{\cL_{MC}\left(\bs_{in}\right)}$ is at most
\begin{align}
     & \quad A\parenv{4 + 2\log Q + Q} + B\left(2t+1\right)\parenv{5 + 2\log Q+\log N}\\
    &+ \bbE\left[\sum_{r=1}^K\Bigl(5\abs{\partial\cC_r}+\abs{\head\left(D_r\right)}  +\abs{\tail\left(D_r\right)}+\abs{F_r}\Bigr)\right] \\ &+ \bbP\parenv{\cE_{\mathrm{col}}^{in}} \cdot \parenv{6N+6+3\log N}+3\log N+6. \label{eq:ELMLubd1_ins}
\end{align}
Condition on the source-symbol lengths, ancestor indices, and insertion locations. Then $K$ and all edit-block lengths are fixed, and each nonempty $D_r$ has i.i.d.\ fair bits, although different edit blocks may be dependent. The head and tail estimates established in the proof of Theorem~\ref{thm:mcd_ub_sub} therefore apply under this conditioning (and are trivial for empty blocks):
\begin{align}
    \bbE\brckt{\abs{\head\left(D_r\right)} } &\le 2^{M+2},\\
    \bbE\brckt{\abs{\tail\left(D_r\right)}}&\le 2^{M+1}+2^{M-1}-M-2 \le 2^{M+2}. \label{eq:mcd_ub_ins_head_tail_exp}
\end{align}
Moreover,
\begin{align}
    \bbE\brckt{\sum_{r=1}^K\abs{\partial\cC_r}}
    &\le 1+\frac{1}{2^{M-1}}\bbE\Bigg[
        \sum_{r=1}^K\Bigl(\abs{\head\left(D_r\right)} +\abs{\tail\left(D_r\right)}+\abs{F_r}\Bigr)\Bigg].
    \label{eq:mcd_ub_ins_partial_ck}
\end{align}
Each conditional head-plus-tail expectation is at most $13\cdot2^{M-1}$ and at most $2^{M+3}$. Averaging over the conditioning and using $K\le B\left(2t+1\right)$ and $\sum_r\abs{F_r}\le Btk$ gives
\begin{align}
    &\bbE\left[\sum_{r=1}^K\Bigl(5\abs{\partial\cC_r}+\abs{\head\left(D_r\right)}  +\abs{\tail\left(D_r\right)}+\abs{F_r}\Bigr)\right]\\
    \le\  &5\left(1+13B\left(2t+1\right)+\frac{Btk}{2^{M-1}}\right) + B\left(2t+1\right)\left(2^{M+3}\right) + Btk. \label{eq:douchkubd_ins}
\end{align}

We next upper bound the probability of $\cE_{\mathrm{col}}^{in}$. Set $n=2^{M-2}$. Since every interior chunk has length at least $2^{M-1}$, a collision in $\cE_{\mathrm{col}}^{in}$ implies equality of two length-$n$ substrings: a substring $\bu$ of a first descendant $Y^{in}_{g\left(a\right)}$ and either (i) a substring $\bv$ at a different position of the same first descendant, (ii) a substring of a first descendant $Y^{in}_{g\left(a'\right)}$ with $a'\ne a$, or (iii) a substring of a boundary chunk that appears before $\bu$.

Condition on the ancestor indices, insertion locations, and source-symbol lengths, and then fix the two candidate starting positions. In case~(i), $Y^{in}_{g\left(a\right)}$ is an i.i.d.\ Bernoulli$\left(1/2\right)$ string under this conditioning. Equality of two distinct length-$n$ substrings therefore has probability $2^{-n}$, including when the substrings overlap. In case~(ii), the two first descendants use independent source symbols and independent insertion bits because $a'\ne a$, so the equality probability is again $2^{-n}$; orient each pair by its later starting position when counting candidates. In case~(iii), a boundary candidate beginning inside $Y^{in}_{g\left(a\right)}$ is already covered by case~(i). Otherwise, no source block preceding $Y^{in}_{g\left(a\right)}$ has ancestor $a$, so the earlier bits are independent of this first descendant. Reveal those earlier bits first, then test the equality constraints from left to right. Each constraint fixes a fresh fair bit of $\bu$, giving probability $2^{-n}$ even when the candidate crosses into the beginning of $Y^{in}_{g\left(a\right)}$. There are at most $AQ$ candidate starting positions for $\bu$. For each such position, there are at most $Q$ other positions in the same first descendant and at most $N$ earlier positions. A union bound therefore gives
\begin{align}
    \bbP\parenv{\cE_{\mathrm{col}}^{in} }
    &\le \frac{AQ\left(Q+N\right)}{2^{2^{M-2}}}.\label{eq:pecomubd_ins}
\end{align}

Thus, by plugging \eqref{eq:douchkubd_ins} and \eqref{eq:pecomubd_ins} into \eqref{eq:ELMLubd1_ins}, we get 
\begin{align}
    \bbE\brckt{\cL_{MC}\left(\bs_{in}\right)} 
    \le & A\parenv{4 + 2\log Q + Q} + B\left(2t+1\right)\parenv{5 + 2\log Q+\log N+2^{M+3}}+5\left(1+13B\left(2t+1\right)+\frac{Btk}{2^{M-1}}\right) \\
    &+ Btk +3\log N+6+2^{-2^{M-2}}\parenv{6N+6+3\log N}AQ\left(Q+N\right)
     , \label{eq:ubdmlE_ins}
\end{align}
subject to the condition that $2^{M-1} \ge \log N$.

Pick 
\begin{align}
    M = \left\lceil\log\left(4\log\parenv{A^2B^{2.5}L^3k^{3.5}}\right)  \right\rceil. \label{eq:mcd_ub_ins_m_choice}
\end{align}
Write $R=A^2B^{2.5}L^3k^{3.5}$. Then $2^{M-2}\ge\log R$, so $2^{-2^{M-2}}\le R^{-1}$, and $2^{M-1} \ge \log N$ for all sufficiently large parameters in the standing regime. Since $t$ is fixed and $k=o\left(L\right)$, the last term in \eqref{eq:ubdmlE_ins} is
\begin{align}
    O\left(\frac{AB^2L^3}{R}\right)=O\left(1\right). \label{eq:mcd_ub_ins_collision_vanishes}
\end{align}
Moreover, the ceiling in \eqref{eq:mcd_ub_ins_m_choice} gives $2^M<8\log R$, and hence $2^{M+3}<64\log R$. Retaining the final-chunk and length-prefix overhead until the final order bound, \eqref{eq:ubdmlE_ins} is further upper bounded by
\begin{align}
    &A\left(4+2\log Q +Q\right) +B\left(2t+1\right)\bigg(5+3\log Q+\log\left(B\right) +64\log\left(A^2B^{2.5}L^3k^{3.5}\right)\bigg)\\
    &\added{+\,}5\biggl(1+13B\left(2t+1\right) +\frac{Btk}{2\log\left(A^2B^{2.5}L^3k^{3.5}\right)}\biggr) +Btk+3\log N+6+O\left(1\right)\\
    = & \ O\parenv{AL} + O\parenv{B\log\left(ABLk\right)} + O\left(Btk\right). \label{eq:mcd_ub_ins_final}
\end{align}
\end{IEEEproof}

\section*{Acknowledgment}
GPT-6 Astra was used to review all proofs and claims for correctness.

\nocite{lou2021asymptotic,motamen2026burst}
\bibliography{bib}

\end{document}